\PassOptionsToPackage{unicode}{hyperref}
\PassOptionsToPackage{hyphens}{url}
\PassOptionsToPackage{dvipsnames,svgnames*,x11names*}{xcolor}
\documentclass[
]{article}
\usepackage{lmodern}
\usepackage{amssymb,amsmath}
\usepackage{ifxetex,ifluatex}
\ifnum 0\ifxetex 1\fi\ifluatex 1\fi=0 
  \usepackage[T1]{fontenc}
  \usepackage[utf8]{inputenc}
  \usepackage{textcomp} 
\else 
  \usepackage{unicode-math}
  \defaultfontfeatures{Scale=MatchLowercase}
  \defaultfontfeatures[\rmfamily]{Ligatures=TeX,Scale=1}
\fi
\IfFileExists{upquote.sty}{\usepackage{upquote}}{}
\IfFileExists{microtype.sty}{
  \usepackage[]{microtype}
  \UseMicrotypeSet[protrusion]{basicmath} 
}{}
\makeatletter
\@ifundefined{KOMAClassName}{
  \IfFileExists{parskip.sty}{%
    \usepackage{parskip}
  }{
    \setlength{\parindent}{0pt}
    \setlength{\parskip}{6pt plus 2pt minus 1pt}}
}{
  \KOMAoptions{parskip=half}}
\makeatother
\usepackage{xcolor}
\IfFileExists{xurl.sty}{\usepackage{xurl}}{} 
\IfFileExists{bookmark.sty}{\usepackage{bookmark}}{\usepackage{hyperref}}
\hypersetup{
  pdftitle={Berlin: A Quantitative View of the Structure of Institutional Scientific Collaborations},
  pdfauthor={Aliakbar Akbaritabar},
  colorlinks=true,
  linkcolor=Maroon,
  filecolor=Maroon,
  citecolor=Blue,
  urlcolor=blue,
  pdfcreator={LaTeX via pandoc}}
\usepackage[margin=1in]{geometry}
\usepackage{longtable,booktabs}
\usepackage{etoolbox}
\makeatletter
\patchcmd\longtable{\par}{\if@noskipsec\mbox{}\fi\par}{}{}
\makeatother
\IfFileExists{footnotehyper.sty}{\usepackage{footnotehyper}}{\usepackage{footnote}}
\makesavenoteenv{longtable}
\usepackage{graphicx,grffile}
\makeatletter
\def\maxwidth{\ifdim\Gin@nat@width>\linewidth\linewidth\else\Gin@nat@width\fi}
\def\maxheight{\ifdim\Gin@nat@height>\textheight\textheight\else\Gin@nat@height\fi}
\makeatother
\setkeys{Gin}{width=\maxwidth,height=\maxheight,keepaspectratio}
\makeatletter
\def\fps@figure{htbp}
\makeatother
\usepackage{pdflscape}
\usepackage{booktabs}
\usepackage{longtable}
\usepackage{array}
\usepackage{multirow}
\usepackage{wrapfig}
\usepackage{float}
\usepackage{colortbl}
\usepackage{pdflscape}
\usepackage{tabu}
\usepackage{threeparttable}
\usepackage{threeparttablex}
\usepackage[normalem]{ulem}
\usepackage{makecell}
\usepackage{xcolor}

\title{Berlin: A Quantitative View of the Structure of Institutional Scientific Collaborations}
\author{Aliakbar Akbaritabar\footnote{German Centre for Higher Education Research and Science Studies (DZHW), Berlin, Germany; \href{mailto:akbaritabar@dzhw.eu}{\nolinkurl{akbaritabar@dzhw.eu}}; \href{mailto:akbaritabar@gmail.com}{\nolinkurl{akbaritabar@gmail.com}}; ORCID = 0000-0003-3828-1533 (Corresponding Author)}}
\date{}

\begin{document}
\maketitle
\begin{abstract}
This paper examines the structure of scientific collaborations in a large European metropolitan area. It aims to identify strategic coalitions among organizations in Berlin as a specific case with high institutional and sectoral diversity. By adopting a global, regional and organization based approach we provide a quantitative, exploratory and macro view of this diversity. We use publications data with at least one organization located in Berlin from 1996-2017. We further investigate four members of the Berlin University Alliance (BUA) through their self-represented research profiles comparing it with empirical results of OECD disciplines. Using a bipartite network modeling framework, we are able to move beyond the uncontested trend towards team science and increasing internationalization. Our results show that BUA members shape the structure of scientific collaborations in the region. However, they are not collaborating cohesively in all disciplines. Larger divides exist in some disciplines e.g., Agricultural Sciences and Humanities. Only Medical and Health Sciences have cohesive intraregional collaborations which signals the success of regional cooperation established in 2003. We explain possible underlying factors shaping the observed trends and sectoral and intra-regional groupings. A major methodological contribution of this paper is evaluating coverage and accuracy of different organization name disambiguation techniques.
\end{abstract}

\textbf{keywords}: Berlin, Internationalization, Co-authorship Network Analysis, Bipartite Community Detection, Disambiguation, Berlin University Alliance

\hypertarget{intro}{%
\section{Introduction}\label{intro}}

Researchers work for academic and non-academic organizations and firms and use the resources from these organizations to carry out scientific work and form scientific collaborations. Coalitions and strategic ties between scientific organizations can be a \emph{cause} and/or an \emph{effect} of the way scientists affiliated to them communicate with each other. An example of the former is the top-down regional, national or organizational policies that support specific types of collaborations (e.g., COST\footnote{\url{https://www.cost.eu/}} initiative to foster scientific networking in Europe). The latter is driven more by the individual motivations of scientists to start bottom-up research projects and obtain funding through inter-organizational collaborations with researchers of other (inter)national organizations (e.g., ERC\footnote{\url{https://erc.europa.eu/}} starting, consolidator or advanced grants). We aim to look at the outcome of scientific collaborations, in form of scientific publications, which is produced through the former process or the latter. By understanding the structure of scientific collaborations between these organizations, we aim to find a proxy to identify possible strategic coalitions among them that in turn could have been inspired by individual researchers.

These strategic coalitions could take different forms and lead to differing set of outputs (Katz \& Martin, \protect\hyperlink{ref-katzWhatResearchCollaboration1997}{1997}; Laudel, \protect\hyperlink{ref-laudelWhatWeMeasure2002a}{2002}). Here we are focused on co-authorship as one of the main forms and scientific publications as the expected output of it. We are aware that co-authorship offers only a reductionist view, but nevertheless it is one of the highly used measures of scientific collaborations.

Moreover, these strategic coalitions can be affected by linguistic (Avdeev, \protect\hyperlink{ref-avdeevInternationalCollaborationHigher2019}{2019}), geographical (Katz, \protect\hyperlink{ref-katzGeographicalProximityScientific1994}{1994}) and regional proximities (Luukkonen, Persson, \& Sivertsen, \protect\hyperlink{ref-luukkonenUnderstandingPatternsInternational1992}{1992}). In an in-depth review, Small \& Adler (\protect\hyperlink{ref-smallRoleSpaceFormation2019}{2019}) presented a diverse array of literature that emphasizes on the effect of \emph{space} in formation of social ties. Scientific organizations are populated by scientists and science is a social enterprise (Fox, \protect\hyperlink{ref-foxPublicationProductivityScientists1983}{1983}). Thus, it is not counterintuitive to consider scientific collaborations as a form of social tie. Formation of these ties are \emph{facilitated} or \emph{hindered} by contextual (Sonnenwald, \protect\hyperlink{ref-sonnenwaldScientificCollaboration2007}{2007}; Small, \protect\hyperlink{ref-smallSomeoneTalk2017}{2017} p 154; Akbaritabar, Casnici, \& Squazzoni, \protect\hyperlink{ref-akbaritabarConundrumResearchProductivity2018}{2018}), social (Smith-Doerr, Alegria, \& Sacco, \protect\hyperlink{ref-smith-doerrHowDiversityMatters2017}{2017}; Akbaritabar \& Squazzoni, \protect\hyperlink{ref-akbaritabarGenderPatternsPublication2020}{2020}) and epistemic preferences of researchers and they can result in denser or instead sparser scientific communities (Akbaritabar et al., \protect\hyperlink{ref-akbaritabarItalianSociologistsCommunity2020}{2020}).

In addition, the increasing trends towards more collaborative work and \emph{team science} is well-known. It is claimed that scientific disciplines, even social sciences, are moving towards more intense collaborations and more internationalization (Wuchty, Jones, \& Uzzi, \protect\hyperlink{ref-wuchtyIncreasingDominanceTeams2007}{2007}; Araújo, Araújo, Moreira, Herrmann, \& Andrade, \protect\hyperlink{ref-araujoGenderDifferencesScientific2017}{2017}). However, studies have highlighted the differences in national or disciplinary contexts in the rate of internationalization (Moed, De Bruin, Nederhof, \& Tijssen, \protect\hyperlink{ref-moedInternationalScientificCooperation1991}{1991}; Babchuk, Keith, \& Peters, \protect\hyperlink{ref-babchukCollaborationSociologyOther1999}{1999}) or differing rates of benefits, in terms of impact, obtained from internationalized collaborations (Glänzel, Schubert, \& Czerwon, \protect\hyperlink{ref-glanzelBibliometricAnalysisInternational1999}{1999}).

It is argued in the literature that scientific and complex economic activities are concentrated in urban and metropolitan areas. In a large-scale study of the USA, Balland et al. (\protect\hyperlink{ref-ballandComplexEconomicActivities2020}{2020}) investigated scientific papers, patents, employment rates and gross domestic product of 353 US metropolitan areas. Their main hypothesis was that disproportionate spatial concentration increases with complexity of productive activities, which was confirmed. They used average number of authors in scientific publications as a proxy of the complexity (due to higher coordination cost of larger scientific teams) and they found that scientific fields with higher complexities tend to have more urban concentration.

In the case of Europe, policies and initiatives are developed aiming at building an ``integrated European Research Area''. Hoekman, Frenken, \& Tijssen (\protect\hyperlink{ref-hoekmanResearchCollaborationDistance2010}{2010}) tested whether this objective is achieved. They concluded that Europe leans towards more integration of previously dominated national contexts. Nevertheless, they reported prevalence of geographically localized co-authorship with high degrees of difference among disciplines in the regional, national or Europe oriented tendencies. They found that some fields, e.g., physical sciences and life sciences are in a more advanced stage of ``Europeanization'' while other fields, e.g., medicine, engineering, social science and humanities present a more nationally oriented scientific collaboration. Furthermore, they concluded that although regional collaborations are still composing high shares of overall collaborations, the effect of territorial and national borders seem to be decreasing over time and Europe is moving towards a more level field of scientific collaboration.

Specific national contexts can present a higher or lower degree of scientific production and internationalized co-authorship. Therefore it is important to take the national context into account along with the continental and regional views. Stahlschmidt, Stephen, \& Hinze (\protect\hyperlink{ref-stahlschmidtPerformanceStructuresGerman2019}{2019}) presented a view of the science system of Germany and provided an updated version of some of their findings in Stephen, Stahlschmidt, \& Hinze (\protect\hyperlink{ref-stephenPerformanceStructuresGerman2020}{2020}). They found that Germany has a stable rate of scientific production similar to that of OECD countries with more established science systems (e.g., the USA, the UK and France), but the growth rate of scientific production of Germany is decreasing. Some countries (e.g., China and India) have higher growth rates of scientific production in recent years. Thus the overall share of Germany from world's scientific publications is decreasing in the recent decades (from 6.3\% in 1995 to 4.3\% in 2018 based on Web of Science (WOS)) (Stephen et al., \protect\hyperlink{ref-stephenPerformanceStructuresGerman2020}{2020}). As presented in Stahlschmidt et al. (\protect\hyperlink{ref-stahlschmidtPerformanceStructuresGerman2019}{2019}), Germany is moving towards higher rates of international collaborations in most scientific disciplines (from 46\% internationalized co-authorships in 2007 to 55\% in 2017 in Scopus and from 47\% in 2007 to 59\% in 2017 in WOS). The USA, the UK, France, Switzerland, Italy, the Netherlands, China, Spain, Austria and Australia are the ten countries with the highest shares of co-authorship with Germany in Scopus. In addition, Aman (\protect\hyperlink{ref-amanHowCollaborationImpacts2016}{2016}) presented evidence of increasing internationalization and also higher rates of citations for inter-organizational and international co-authorship for the German science system in WOS from 2007 to 2012.

There is also studies specifically focused on the Berlin metropolitan region. Rammer, Kinne, \& Blind (\protect\hyperlink{ref-rammerKnowledgeProximityFirm2020}{2020}) provide a fine-grained view of the case of Berlin and how a form of selective spatial proximity exists between knowledge producing institutions (e.g., universities) and knowledge demanding institutions (e.g., innovative companies and firms). They used the first wave of panel data curated through Berlin Innovation Panel which surveys enterprises with five or more employees in manufacturing and knowledge-intensive services located in Berlin. They reported a micro-geographic scope where innovative firms are surrounded by same-sector firms and they were located closer to universities and research institutes that can signal a selective process. They also found that change in innovative activities of a given firm affects other firms in their vicinity. It is necessary to note that the sample analyzed in their study was composed of 80\% small firms and 70\% service sector which can affect their conclusions.

Abbasiharofteh \& Broekel (\protect\hyperlink{ref-abbasiharoftehStillShadowWall2020}{2020}) explored the biotechnology field in Berlin metropolitan region. They investigated co-patenting, co-authorship and joint R\&D collaborations and provided an exhaustive view of the temporal changes in the scientific landscape of the region in the specific case of biotechnology. They concluded that eastern and western organizations within Berlin are still not cohesively collaborating with each other. The ``shadow'' of the Berlin wall still influences the scientific collaborations of the region.

Organizational, regional and continental agreements and strategic coalitions are being developed to support higher rates of scientific collaboration among the actors in these contexts. A specific example is \emph{Berlin University Alliance} (BUA)\footnote{\url{https://www.berlin-university-alliance.de/en/about/index.html}}. BUA was founded in February 2018 between the three main universities and one university hospital located in the Berlin metropolitan region i.e., \emph{Freie Universität Berlin} (FU), \emph{Humboldt-Universität zu Berlin} (HU), \emph{Technische Universität Berlin} (TU) and \emph{Charité -- Universitätsmedizin Berlin} (CH) (Berlin University Alliance, \protect\hyperlink{ref-berlinuniversityallianceGemeinsamImVerbund2018}{2018}, \protect\hyperlink{ref-berlinuniversityallianceBerlinUniversityAlliance2019}{2019}). BUA claims to be established based on a long lasting record of intra-regional collaborations between these institutions. The interaction between these institutions have started from an era of institutional isolation after the fall of Berlin wall during which these institutes needed to define and empower their unique identities. Afterwards, first forms of cooperation between these institutes emerged which lead in 2003 to establishment of a shared medical faculty between HU and FU to be located in Charité and allow higher collaborations in medical and natural sciences. There are examples of competition, mutual definition of exclusive research areas and graduate programs versus close cooperation among BUA members in the past three decades. These are highlighted in the BUA proposal as strengths of the region. Although the four major institutions of the region have their own unique identity and research profiles (further detail will follow in \protect\hyperlink{datamethods}{Data and Methods} section), nevertheless, they aim at fostering previous collaboration experiences in a new organizational form, i.e., BUA. We aim to control whether these four institutions have a distinctive position in the structure of scientific collaborations formed in the Berlin metropolitan region. Thus, we add specific measures to control whether BUA members form cohesive collaboration ties among themselves. As the research profiles presented later in text advocates, these institutions have overlapping disciplinary focuses. We aim to investigate whether they have prevailing roles in these overlapping disciplines or whether we can find signs of strategic coalitions among them i.e., cohesive co-authorship communities.

Network analysis can be used to identify the presence of communities in co-authorship networks (Palla, Barabási, \& Vicsek, \protect\hyperlink{ref-pallaQuantifyingSocialGroup2007}{2007}; Leone Sciabolazza, Vacca, Kennelly Okraku, \& McCarty, \protect\hyperlink{ref-leonesciabolazzaDetectingAnalyzingResearch2017}{2017}; Akbaritabar et al., \protect\hyperlink{ref-akbaritabarItalianSociologistsCommunity2020}{2020}). Quantitative models are used to examine if collaboration patterns persist \emph{between} or \emph{within} denser areas of the network and in form of specific communities. Looking at the composition of these communities and identifying potential factors contributing to their cohesion helps to explain groupings in scientific collaborations.

In lower levels than continental, national or regional frameworks, scientific organizations themselves could have strategic plans to define their overarching identities and main research focus. This might inspire researchers in a certain organization to prioritize research in specific disciplines and areas (Blume, Bunders, Leydesdorff, \& Whitley, \protect\hyperlink{ref-blumeSocialDirectionPublic1987}{1987}) to signal allegiance with the organization's designed identity which in turn could penalize selection of innovative research themes (Rijcke, Wouters, Rushforth, Franssen, \& Hammarfelt, \protect\hyperlink{ref-rijckeEvaluationPracticesEffects2016}{2016}). Collaboration can be affected by the goals set out by funding agencies (Nederhof, \protect\hyperlink{ref-nederhofBibliometricMonitoringResearch2006}{2006}; Wagner, Park, \& Leydesdorff, \protect\hyperlink{ref-wagnerContinuingGrowthGlobal2015}{2015}). Furthermore, it can be affected by the type of organization (i.e., sector) which partially determines the type of research of an organization or expected outcomes of it.

In addition to the themes discussed above, the type of data employed to answer the research questions could have a large effect on the identified trends. Bibliometric databases are not perfect and any given one could be prone to specific errors. In terms of coverage, different databases have certain policies to define what should be indexed. This affects results of macro studies depending on the database employed, subset of scientific publications used, document types analyzed and the level of aggregation and normalization applied (Stahlschmidt et al., \protect\hyperlink{ref-stahlschmidtPerformanceStructuresGerman2019}{2019}; Stephen et al., \protect\hyperlink{ref-stephenPerformanceStructuresGerman2020}{2020}). In terms of cleanness of the data, there is a strong need for disambiguation of scientific entity (e.g., authors, organizations) names which might bias the quality of results (Aman, \protect\hyperlink{ref-amanDoesScopusAuthor2018}{2018}; Donner, Rimmert, \& van Eck, \protect\hyperlink{ref-donnerComparingInstitutionallevelBibliometric2019}{2019}; D'Angelo \& van Eck, \protect\hyperlink{ref-dangeloCollectingLargescalePublication2020}{2020}). Thus, one of our methodological goals is to introduce organization name disambiguation techniques to match scientific organization names with publicly available databases (e.g., Wikidata and Global Research Identifier Database (GRID)) and evaluate the reliability of the results in comparison to established techniques.

Based on this literature, we intend to explore the interplay between different contextual variables and space to investigate the structure of scientific collaborations in the Berlin metropolitan region. Thus, on the one hand, we focus on Berlin as a regional hub or context that can inspire specific organizational arrangements, e.g., BUA. Furthermore, institutions located in Germany and presumably true for institutions in Berlin, collaborate with many institutions worldwide (Stahlschmidt et al., \protect\hyperlink{ref-stahlschmidtPerformanceStructuresGerman2019}{2019}), hence, we expect to see a high degree of internationalized collaborations. On the other hand, the Berlin metropolitan region hosts a diverse group of national and international researchers affiliated to scientific organizations (Berlin University Alliance, \protect\hyperlink{ref-berlinuniversityallianceBerlinUniversityAlliance2019}{2019} p 13), thus, we expect to see a wide array of research outputs in multiple disciplines and disciplinary specific trends of collaboration. Therefore, we formulate and investigate the following macro, quantitative and exploratory research questions:

\begin{itemize}
\item
  \textbf{RQ1}: How \emph{collaborative} and \emph{internationalized} is the scientific landscape of the Berlin metropolitan region?
\item
  \textbf{RQ2}: Are there \emph{disciplinary} differences in the rate of collaborative and internationalized scientific work?
\item
  \textbf{RQ3}: How \emph{regionally} or \emph{continentally} oriented is scientific collaboration in the Berlin metropolitan region?
\item
  \textbf{RQ4}: How \emph{sector} oriented is scientific collaboration in the Berlin metropolitan region?
\item
  \textbf{RQ5}: How does the \emph{diversity} and level of development of national or regional science systems worldwide affect the collaborations with the Berlin metropolitan region?
\item
  \textbf{RQ6}: Is there evidence of strategic coalitions, disciplinary, regional or organizational \emph{agreements} in the structure of scientific collaborations in the Berlin metropolitan region?
\item
  \textbf{RQ7}: Are there specific disciplinary, sectoral, national or continental \emph{cohesive subgroups} driving the scientific collaborations in the Berlin metropolitan region?
\item
  \textbf{RQ8}: How influential is the \emph{disambiguation} of organization names on the structure of scientific collaborations?
\end{itemize}

Contributions of the current paper is fourfold: 1) we focus on scientific output of the Berlin metropolitan region and trace the share of collaborative works and identify the share of international collaborations. We separate Berlin, Germany, Europe and continental regions worldwide to investigate possible groupings and we intend to move beyond the descriptive and macro view, which advocates for increasing internationalization, by investigating the structure of scientific collaborations in a multi-level framework. 2) we cover six major OECD scientific disciplines and provide a comparative view of the specificities of these disciplines and we include a sectoral view based on the type of organizations. 3) we develop and use multiple organization name disambiguation techniques and compare their efficiency, coverage and accuracy and 4) we employ a bipartite network modeling and community detection approach and present how it can be useful in co-authorship network analysis and identification of denser groups collaborating preferably among themselves. Since investigation on the level of entire organization can cause a high rate of interconnectivity in the network (due to aggregation to organization level and overlooking individual researcher or research group borders), as described in \protect\hyperlink{datamethods}{Data and Methods} Section, our bipartite modeling approach takes specificities in \emph{single} publication level into account and provides a more fine-grained view of groups of organizations contributing to each publication.

The structure of the paper is as follows: \protect\hyperlink{datamethods}{Data and Methods} section presents our data sources and modeling strategy. \protect\hyperlink{results}{Results} section presents our findings which is followed by \protect\hyperlink{conclusions}{Discussion} section.

\hypertarget{datamethods}{%
\section{Data and methods}\label{datamethods}}

We use Scopus 2018 data from the German Bibliometrics Competence Center (KB)\footnote{Kompetenzzentrum Bibliometrie (KB), \url{http://bibliometrie.info}}. We extract \emph{article}, \emph{review} and \emph{conference proceedings} documents published from beginning of the database in 1996 till end of 2017. To delineate \emph{Berlin} metropolitan region and to identify the scientific collaborations occurred in the region, we select only publications that have at least one authoring organization located in Germany and Berlin. Thus, co-authorship here includes Berlin organizations and their collaborators worldwide. Our level of analysis are \emph{scientific organizations} (i.e., each affiliation address mentioned in a publication that can be academic or non-academic organizations or firms where researchers affiliate) and we do not investigate lower levels e.g., authors, since our goal is to identify structure of scientific collaborations among organizations.

Our data include different meta-data for each publication such as \emph{publication year}, \emph{title}, \emph{affiliation addresses}, \emph{scientific discipline}, \emph{journal name} and \emph{document type}. We include \emph{conference proceedings} in addition to articles and reviews since there are technical universities in the sample for which this type of document is considered influential. We use a mapping of publications to OECD scientific disciplines based on Scopus ASJC\footnote{All Science Journal Classification}. We compare the aggregate data with trends of different OECD scientific disciplines i.e., \emph{Agricultural Sciences} (AS), \emph{Engineering Technology} (ET), \emph{Natural Sciences} (NS), \emph{Medical and Health Sciences} (MHS), \emph{Humanities} (H) and \emph{Social Sciences} (SS). Please note that some publications might be assigned to multiple disciplines. In the aggregate analysis, we use the first assignment of each publication, but in a single discipline view, we take publications with any assignment in the given discipline, thus, interdisciplinary publications are covered separately in all their assigned disciplines.

As described earlier, scientific organizations set goals and define strategic paths to ensure a unique research profile and identity. In order to have a better understanding of how BUA members introduce their own research goals and main areas, we use their self-representations in Berlin University Alliance (\protect\hyperlink{ref-berlinuniversityallianceBerlinUniversityAlliance2019}{2019}) and Berlin University Alliance (\protect\hyperlink{ref-berlinuniversityallianceGemeinsamImVerbund2018}{2018}). We expect to observe prevailing roles of these institutions in structure of scientific collaborations of the disciplines closer to their areas of focus. \textbf{FU}: ``Biomedical Foundations'', ``Complex Systems'', ``Cultural Dynamics'', ``Educational Processes and Results'', ``Health and Quality of Life'', ``Human-Environmental Interactions'', ``In-Security and Security Research'', ``Materials Research'' and ``Transregional Relations''. \textbf{HU}: ``Application-Oriented Mathematics'', ``Image Sciences'', ``Integrative Life Sciences'', ``Integrative Natural Sciences'', ``Research on Law and Society'', ``Study of Ancient Civilizations'' and ``Sustainability Research''. \textbf{CH}: ``Cardiovascular Research \& Metabolism'', ``Infection, Immunology \& Inflammation'', ``Neuroscience'', ``Oncology'', ``Rare Disease \& Genetics'' and ``Regenerative Therapies''. \textbf{TU}: ``Materials, Design and Manufacturing'', ``Digital Transformation'', ``Energy Systems, Mobility and Sustainable Resources'', ``Urban and Environmental Systems'', ``Optic and Photonic Systems'' and ``Education and Human Health''. Except CH which has a focus on MHS and NS, the other three institutions are active in areas close to major OECD disciplines.

\hypertarget{organization-name-disambiguation}{%
\subsection{Organization name disambiguation}\label{organization-name-disambiguation}}

The data delivered by Scopus is not perfect. It is prone to error and there is a strong need for \emph{disambiguation} of organization names (Donner et al., \protect\hyperlink{ref-donnerComparingInstitutionallevelBibliometric2019}{2019}). Without disambiguation, co-authorship networks constructed will have multiple representations of the same actor and an artificially higher level of (dis)connectivity.

\begin{figure}

{\centering \includegraphics[width=1\linewidth,]{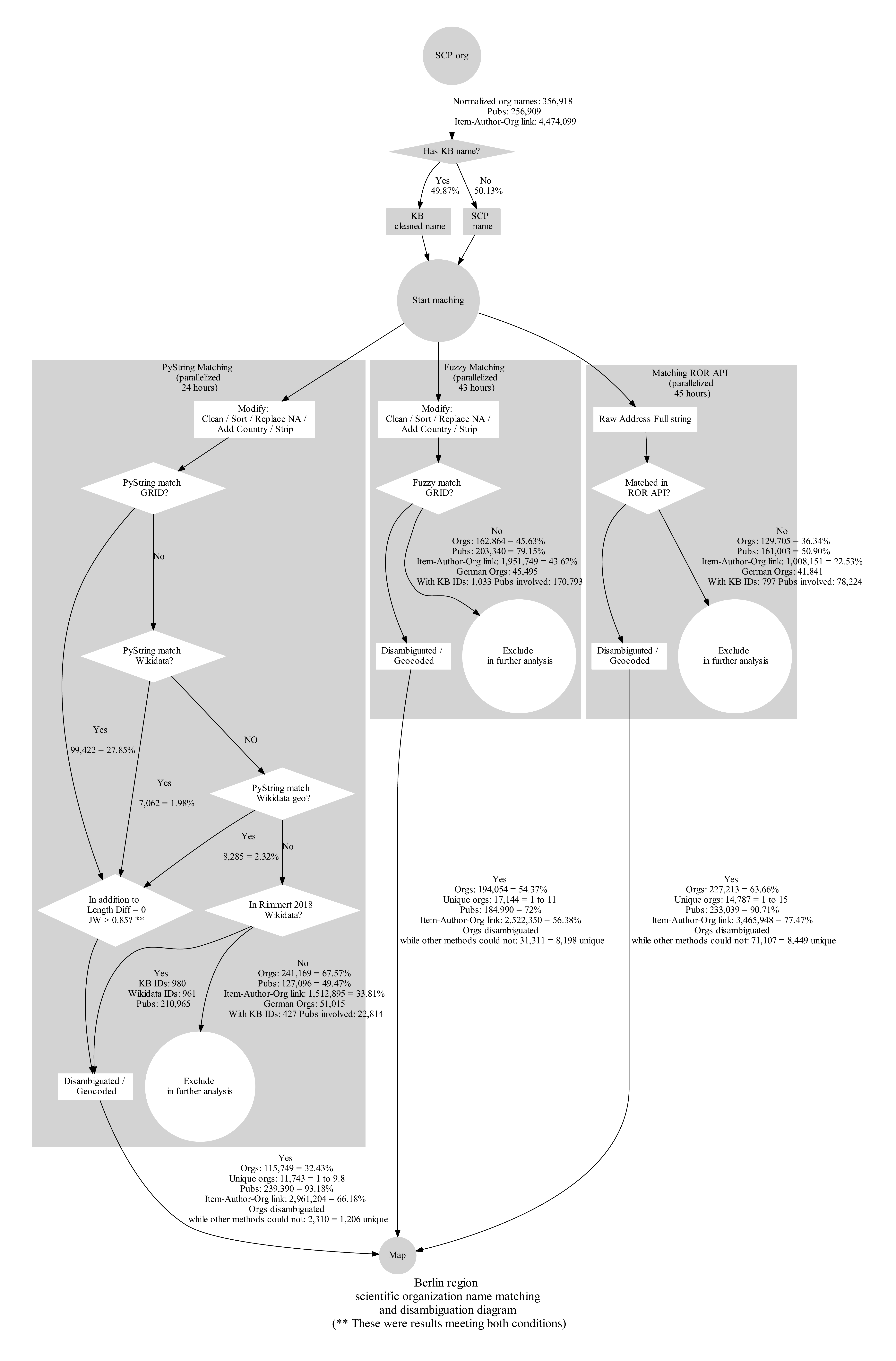} 

}

\caption{Organization name disambiguation techniques and comparison of coverage and accuracy}\label{fig:disambiguation-logic-and-results}
\end{figure}

We developed two disambiguation techniques (i.e., \emph{PyString} and \emph{Fuzzy} matching) and compared their results with a previously established technique (i.e., Research Organization Registry (ROR)\footnote{\url{https://ror.org/about}}), as depicted in figure \ref{fig:disambiguation-logic-and-results}. ROR uses data from Global Research Identifier Database (GRID\footnote{\url{https://www.grid.ac/pages/policies}}) prepared by Digital Science\footnote{\url{https://www.digital-science.com/}}, ISNI\footnote{International Standard Name Identifier, \url{https://isni.org/}}, Crossref and Wikidata\footnote{\url{https://www.wikidata.org}}.

In \emph{PyString} matching (gray shaded area on the left of figure \ref{fig:disambiguation-logic-and-results}), we standardize organization names and perform a match with GRID (snapshot of February 17, 2019). We use only the largest entity that Kompetenzzentrum Bibliometrie extracts from the Scopus using the first part of affiliation string before the first comma. To match, we used \emph{string} comparison methods in Python (for simplicity we call it \emph{PyString}) which matches whole and subsets of the text strings but does not account for change in the order of words in organization names. To remove the effect of order of organization name parts, we split the names based on space (i.e., words) and reorder them alphabetically for both Scopus and GRID entities. We then add country names to the end of strings to allow higher precision of matching and reduce the effect of organizational homonyms\footnote{As an example, from this address string delivered by Scopus, ``Freie Universität Berlin, Department of education and psychology, DEU'', KB extracts ``Freie Universität Berlin'' as the first part which we use for PyString and Fuzzy matching processes and after removing alphanumeric symbols, lowercasing and reordering alphabetically, we add the 3-digit ISO country code to the end (e.g., ``berlin freie universität DEU'').}. For those still non-matched, we perform another PyString match with scientific organizations in Wikidata and for the still missing ones, with Wikidata entities which have geographical coordinates. We limit the results to the most promising ones using Jaro Winkler distance of more than 0.85 between the two matched names. We do this after the PyString match is done, as a control of reliability. We chose this threshold based on manual evaluation of match results to have the highest accuracy. At the end and to complement the results of PyString procedure, we search organization names in an in-house database which is previously developed by comparing organization names to Wikidata entities by Rimmert (\protect\hyperlink{ref-rimmertInstitutionalDisambiguationFurther2018}{2018})\footnote{This in-house data is only accessible through KB infrastructure.}.

In parallel, we compare organization names with GRID by \emph{Fuzzy} matching the names. This method takes differing word order and subsets of the name into account (we standardize the names as before and add country). For Fuzzy text matching, we use the \emph{FuzzyWuzzy}\footnote{\url{https://github.com/seatgeek/fuzzywuzzy}} library in Python. Using \emph{fuzz.ratio} as scorer, we set a threshold of 80 percent which was chosen based on empirical evaluation on some exemplar cases and proved to give a reliable accuracy (gray shaded area in center of figure \ref{fig:disambiguation-logic-and-results}).

In a third attempt, instead of the main organization names used in previous procedures, we used the complete string of affiliation addresses delivered by Scopus to disambiguate it with Research Organization Registry (ROR) API (see previous footnote for an example). We obtain further information (i.e., country, geographical coordinates (longitude and latitude) of the main address and type of organization as \emph{education}, \emph{non-profit}\footnote{Organization that uses its surplus revenue to achieve its goals. Includes charities and other non-government research funding bodies. Example, the Max Planck Society (grid.4372.2)}, \emph{company}, \emph{government}, \emph{health-care}, \emph{facility}\footnote{A building or facility dedicated to research of a specific area, usually contains specialized equipment. Includes telescopes, observatories and particle accelerators. Example: member institutes of the Max Planck Society (e.g., Max Planck Institute for Demographic Research, grid.419511.9)}, \emph{archive}\footnote{Repository of documents, artifacts, or specimens. Includes libraries and museums that are not part of a university. Example, New York Public Library (grid.429888.7)} and other). We used the ROR snapshot from November 7\textsuperscript{th} 2019. This disambiguation takes different name spellings and misspelled words, acronyms and multiple languages into account. In order not to face API request limits, we have set up a local instance of ROR API (see gray shaded area on the right of figure \ref{fig:disambiguation-logic-and-results}).

To clarify the importance and effect of the disambiguation, we define different scenarios based on non-disambiguated and disambiguated data. We present the results and implications of them in construction of the co-authorship networks. However, for our disciplinary, geographical and sector analysis and on the basis of the accuracy and coverage results presented later, we use the third disambiguation technique described above (i.e., ROR). It is important to note that in disambiguated versions of the data, we include only publications for which all contributing organizations are disambiguated and we exclude those publications with one or more non-disambiguated co-authoring organizations.

\hypertarget{bipartite-network-modeling}{%
\subsection{Bipartite network modeling}\label{bipartite-network-modeling}}

We construct bipartite co-authorship networks (Breiger, \protect\hyperlink{ref-breigerDualityPersonsGroups1974}{1974}) using ties between publications and \emph{organizations} (Katz \& Martin, \protect\hyperlink{ref-katzWhatResearchCollaboration1997}{1997}). We treat each single publication as an event where organizations interact to produce an academic text (Biancani \& McFarland, \protect\hyperlink{ref-biancaniSocialNetworksResearch2013}{2013}). Studies on co-authorship networks usually use a one-mode projection of these bipartite networks (Newman, \protect\hyperlink{ref-newmanScientificCollaborationNetworks2001}{2001}\protect\hyperlink{ref-newmanScientificCollaborationNetworks2001}{a}, \protect\hyperlink{ref-newmanScientificCollaborationNetworks2001a}{2001}\protect\hyperlink{ref-newmanScientificCollaborationNetworks2001a}{b}). The problem with this projection is twofold. Different structures in two-mode networks are projected to the same one-mode structure which causes an information loss about the underlying structure. Second, the one-mode projection can present an artificially higher density and connectivity due to publications with high number of authors which project to maximally connected cliques. By adopting methods specifically developed for \emph{bipartite networks} we are able to resolve the shortcomings.

To identify possible geographical, disciplinary and/or sector based coalitions between scientific organizations, we extract the largest connected component of the network, i.e., giant component, and investigate it further. Our aim is to see if there are cohesive subgroups of organizations preferably collaborating \emph{among} themselves. We investigate the potential underlying factors behind these groupings.

To identify communities of co-authorship, we use \emph{bipartite community detection} by \emph{Constant Potts model} (CPM). CPM is a specific version of Potts model (Reichardt \& Bornholdt, \protect\hyperlink{ref-reichardtDetectingFuzzyCommunity2004}{2004}) proposed by Traag, Van Dooren, \& Nesterov (\protect\hyperlink{ref-traagNarrowScopeResolutionlimitfree2011}{2011}) as a \emph{resolution-limit-free} method. It resolves the resolution limit problem in modularity (Newman, \protect\hyperlink{ref-newmanDetectingCommunityStructure2004}{2004}) which can obstruct detection of small communities in large networks (Traag, Waltman, \& van Eck, \protect\hyperlink{ref-traagLouvainLeidenGuaranteeing2019}{2019}). We use the implementation in the \emph{Leidenalg}\footnote{\url{https://github.com/vtraag/leidenalg}} library in Python. Community detection emphasizes the importance of links \emph{within} communities rather than those \emph{between} them. CPM uses a resolution parameter \(\gamma\) (i.e., ``\emph{constant}'' in the name), leading to communities such that the link density between the communities (external density) is lower than \(\gamma\) and the link density within communities (internal density) is more than \(\gamma\). We set different resolution parameters in case of aggregate data (ROR = \(3 \times 10^{-3}\)) and scientific disciplines (AS = \(7 \times 10^{-4}\), ET = \(6 \times 10^{-3}\), NS = \(6 \times 10^{-3}\), MHS = \(5 \times 10^{-3}\), H = \(4 \times 10^{-4}\) and SS = \(6 \times 10^{-3}\)). We chose these parameters after exploration of the number of communities detected in contrast to the number of organizations and publications included in each bipartite community to arrive at a rather consistent distribution.

\hypertarget{results}{%
\section{Results}\label{results}}

\hypertarget{implications-of-organization-name-disambiguation}{%
\subsection{Implications of organization name disambiguation}\label{implications-of-organization-name-disambiguation}}

Figure \ref{fig:disambiguation-logic-and-results} presents different disambiguation techniques used and the coverage of item(publication)-author-organization links (\emph{PyString} 66.18\%, \emph{Fuzzy} 56.38\% and \emph{ROR} 77.47\%). We present author level counts of links here since different authors from the same institution might mention different affiliation addresses by including department names or they might report erroneous addresses. However, in building organization level co-authorship networks, we exclude repeated organization-publication links. Each technique successfully disambiguates a set of unique organization names which other techniques are unable to disambiguate (PyString 1,206, Fuzzy 8,198 and ROR 8,449). Note that not all organizations involved in authoring a publication are successfully disambiguated (PyString 115,749 (32.43\%) organizations from 239,390 (93.18\%) publications, Fuzzy 194,054 (54.37\%) organizations from 184,990 (72\%) publications and ROR 227,213 (63.66\%) organizations from 233,039 (90.71\%) publications). We only include publications for which all contributing organizations are successfully disambiguated and this decreases our coverage to 129,813 (51\%) publications in PyString, 53,569 (21\%) in Fuzzy and 126,130 (49\%) in ROR. Table \ref{tab:description-different-disambiguation} compares the networks constructed using non-disambiguated data with the output of different disambiguation techniques. We observed a high rate of dis-connectivity in the non-disambiguated network (10,269 components) while this was extremely reduced through disambiguation techniques (i.e., to 66 in PyString, 159 in Fuzzy and 100 in ROR). The share of nodes in the giant component which was initially high in the non-disambiguated network (95\%) further increased and covered close to 99\% in all cases (numbers in the table are rounded up). In all these cases, disambiguation shows that many unique organization names delivered by Scopus need to be merged due to spelling error and name order changes which can affect the networks constructed to a high degree (see De Stefano, Fuccella, Vitale, \& Zaccarin (\protect\hyperlink{ref-destefanoUseDifferentData2013}{2013}) for a discussion of possible effects). In PyString, the ratio of disambiguated unique organizations to non-disambiguated ones were 1 to 9.8 (in Fuzzy 1 to 11 and in ROR 1 to 15). This proves the high influence disambiguation could have on the results (\textbf{RQ8}). Table \ref{tab:description-different-discipline-networks} presents the networks in different OECD scientific disciplines using ROR results. Note that the results which follow are based on the 49\% of publications that were successfully disambiguated by ROR technique. Each of the disciplines presented in Table \ref{tab:description-different-discipline-networks} covers a different share of connected components observed in the aggregate data ranging from 13 components in Agricultural Sciences (AS) to 68 in Social Sciences (SS). Natural Sciences (NS) has both the highest number of publications and organizations while Humanities (H) has the smallest number of publications and organizations.

\begin{table}

\caption{\label{tab:description-different-disambiguation}Berlin organizations co-authorship networks using non-disambiguated and disambiguated data (G = giant component)}
\centering
\resizebox{\linewidth}{!}{
\begin{tabular}[t]{l|l|l|l|l}
\hline
Metrics                               &   Non disambiguated &   PyString &   Fuzzy &     ROR\\
\hline
N. of connected components & 10,269 & 66 & 159 & 100\\
\hline
N. of biparitite nodes & 613,827 & 135,057 & 58,547 & 133,387\\
\hline
N. of biparitite edges & 1,083,775 & 246,704 & 89,199 & 246,472\\
\hline
\% of biparitite nodes in G & 95 & 100 & 99 & 100\\
\hline
\% of biparitite edges in G & 98 & 100 & 100 & 100\\
\hline
N. of organizations & 356,918 & 5,244 & 4,978 & 7,257\\
\hline
N. of organizations in G & 337,755 & 5,176 & 4,809 & 7,153\\
\hline
N. of publications (\%) & 256,909 & 129,813 (51\%) & 53,569 (21\%) & 126,130 (49\%)\\
\hline
N. of publications in G & 245,203 & 129,657 & 53,248 & 125,949\\
\hline
\end{tabular}}
\end{table}

\begin{table}

\caption{\label{tab:description-different-discipline-networks}Berlin organizations co-authorship networks in different OECD scientific disciplines (G = giant component, ROR organization name disambiguation)}
\centering
\resizebox{\linewidth}{!}{
\begin{tabular}[t]{l|r|r|r|r|r|r}
\hline
Metrics                               &     AS &     ET &     H &    MHS &      NS &     SS\\
\hline
N. of connected components & 13 & 56 & 34 & 48 & 55 & 68\\
\hline
N. of biparitite nodes & 8,528 & 33,930 & 4,835 & 46,842 & 89,032 & 16,045\\
\hline
N. of biparitite edges & 13,822 & 57,671 & 6,195 & 84,945 & 170,334 & 25,336\\
\hline
\% of biparitite nodes in G & 100 & 100 & 98 & 100 & 100 & 99\\
\hline
\% of biparitite edges in G & 100 & 100 & 99 & 100 & 100 & 100\\
\hline
N. of organizations & 1,687 & 2,991 & 798 & 3,843 & 5,970 & 2,091\\
\hline
N. of organizations in G & 1,668 & 2,933 & 763 & 3,792 & 5,910 & 2,012\\
\hline
N. of publications & 6,841 & 30,939 & 4,037 & 42,999 & 83,062 & 13,954\\
\hline
N. of publications in G & 6,828 & 30,851 & 3,988 & 42,913 & 82,989 & 13,849\\
\hline
\end{tabular}}
\end{table}

\hypertarget{macro-view-of-scientific-output-of-berlin-metropolitan-region}{%
\subsection{Macro view of scientific output of Berlin metropolitan region}\label{macro-view-of-scientific-output-of-berlin-metropolitan-region}}

\begin{figure}

{\centering \includegraphics[width=1\linewidth,]{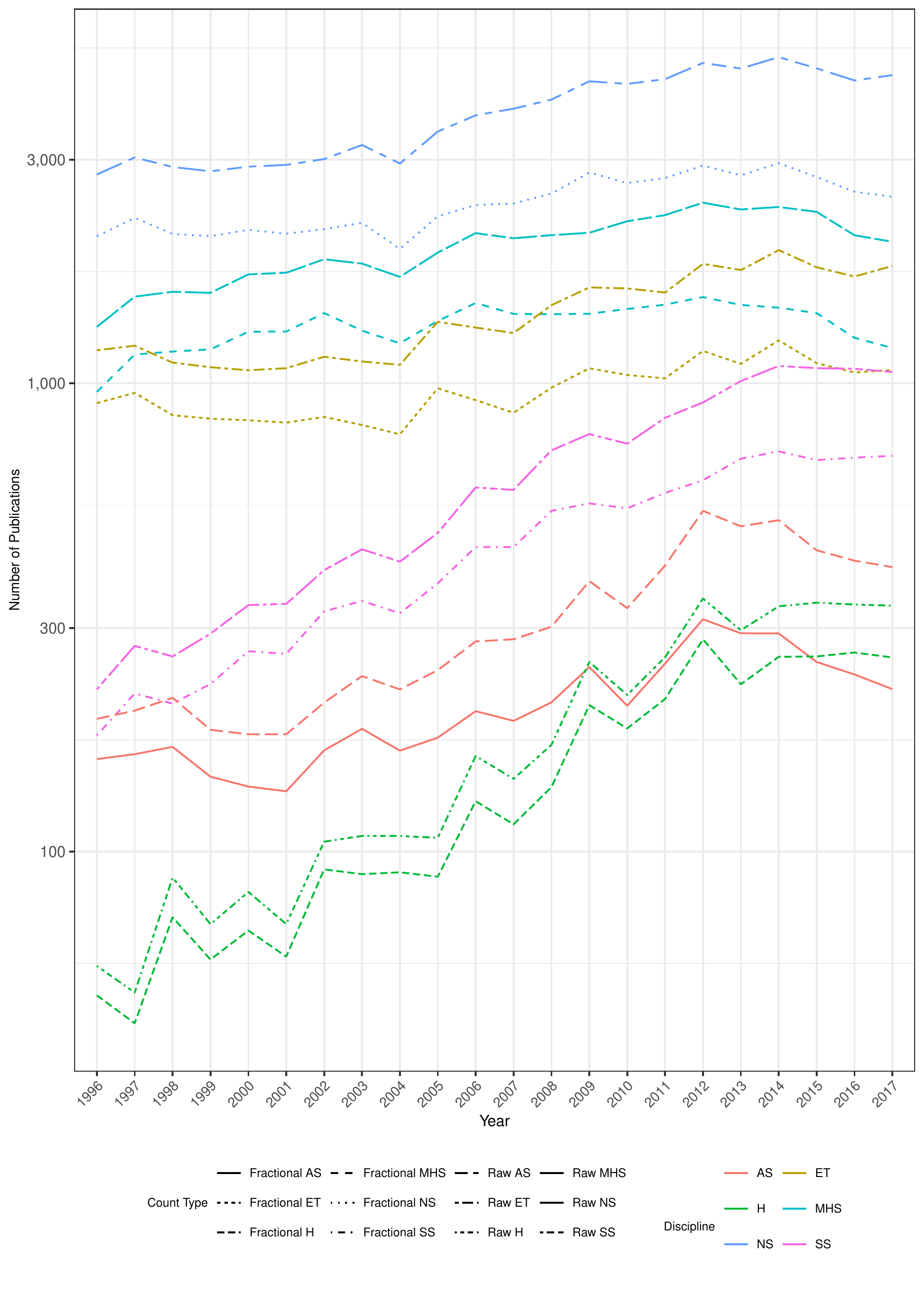} 

}

\caption{Raw and fractional count of Berlin publications by OECD disciplines (1996-2017, Scopus, fractional count based on organizations, Y on log scale)}\label{fig:range-of-publications}
\end{figure}

\begin{figure}

{\centering \includegraphics[width=1\linewidth,]{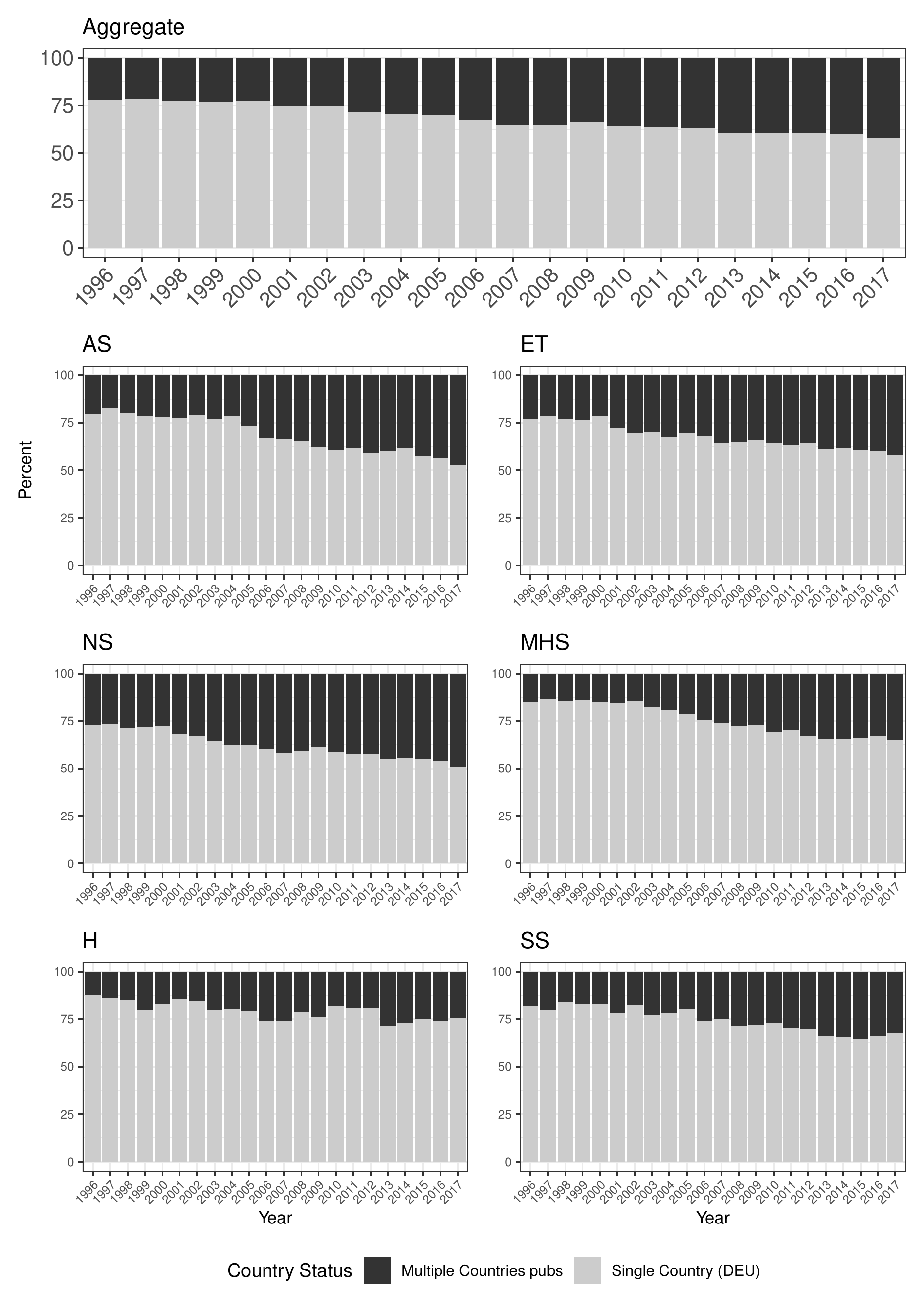} 

}

\caption{Share of intra-Germany versus multiple country co-authorship, (Top) aggregate (Bottom) different disciplines (1996-2017, Scopus)}\label{fig:single-multiple-country-publications-disciplines}
\end{figure}

Figure \ref{fig:range-of-publications} presents the raw and fractional count of publications among different OECD disciplines. Note that it is based on publications which have at least one collaborator from Berlin metropolitan region and for organizations that were successfully disambiguated with ROR technique. Nevertheless, the trends are in-line with what Stephen et al. (\protect\hyperlink{ref-stephenPerformanceStructuresGerman2020}{2020}) and Stahlschmidt et al. (\protect\hyperlink{ref-stahlschmidtPerformanceStructuresGerman2019}{2019}) reported for Germany. Some disciplines show higher rates of collaborative work (e.g., see case of NS, blue lines, first and second from top) which is evident in the gap between the lines presenting their raw and fractional counts. In contrast, some disciplines that are traditionally known to be less collaborative (Leahey, \protect\hyperlink{ref-leaheySoleInvestigatorTeam2016}{2016}) present a smaller gap on the plot (e.g., Humanities). Since the Y axis is on log 10 scale, the figure shows growth rates in raw and fractional count of publications which is tripled in case of Humanities and Social Sciences. But it can be due to higher coverage of Scopus in recent years and not merely increase of publications (see Stahlschmidt et al. (\protect\hyperlink{ref-stahlschmidtPerformanceStructuresGerman2019}{2019}) for a discussion). To investigate internationalization of publications, figure \ref{fig:single-multiple-country-publications-disciplines} presents the \emph{single} (intra-DEU) versus \emph{multiple} country publications. Despite the fact that it presents only the 49\% of publications which were disambiguated, but the trends observed are in line with the case of German science system reported in Stahlschmidt et al. (\protect\hyperlink{ref-stahlschmidtPerformanceStructuresGerman2019}{2019}). It is clear that some disciplines have already reached close to 50\% share of \emph{internationalization} (i.e., NS) driving the aggregate trend of increasing internationalization observed on top panel of the figure. However, there are other disciplines with still lower than 25\% share of international collaborations (i.e., H and SS). In all these disciplines an increasing trend towards further internationalization is evident (see increasing length of black bars in the figure) except H and SS that do not present a clear increasing trend and in some years the rate of internationalization decreases. This answers our \textbf{RQ1} and \textbf{RQ2} signaling a high disciplinary difference in rate of collaborative work and internationalization.

\begin{figure}

{\centering \includegraphics[width=1\linewidth,]{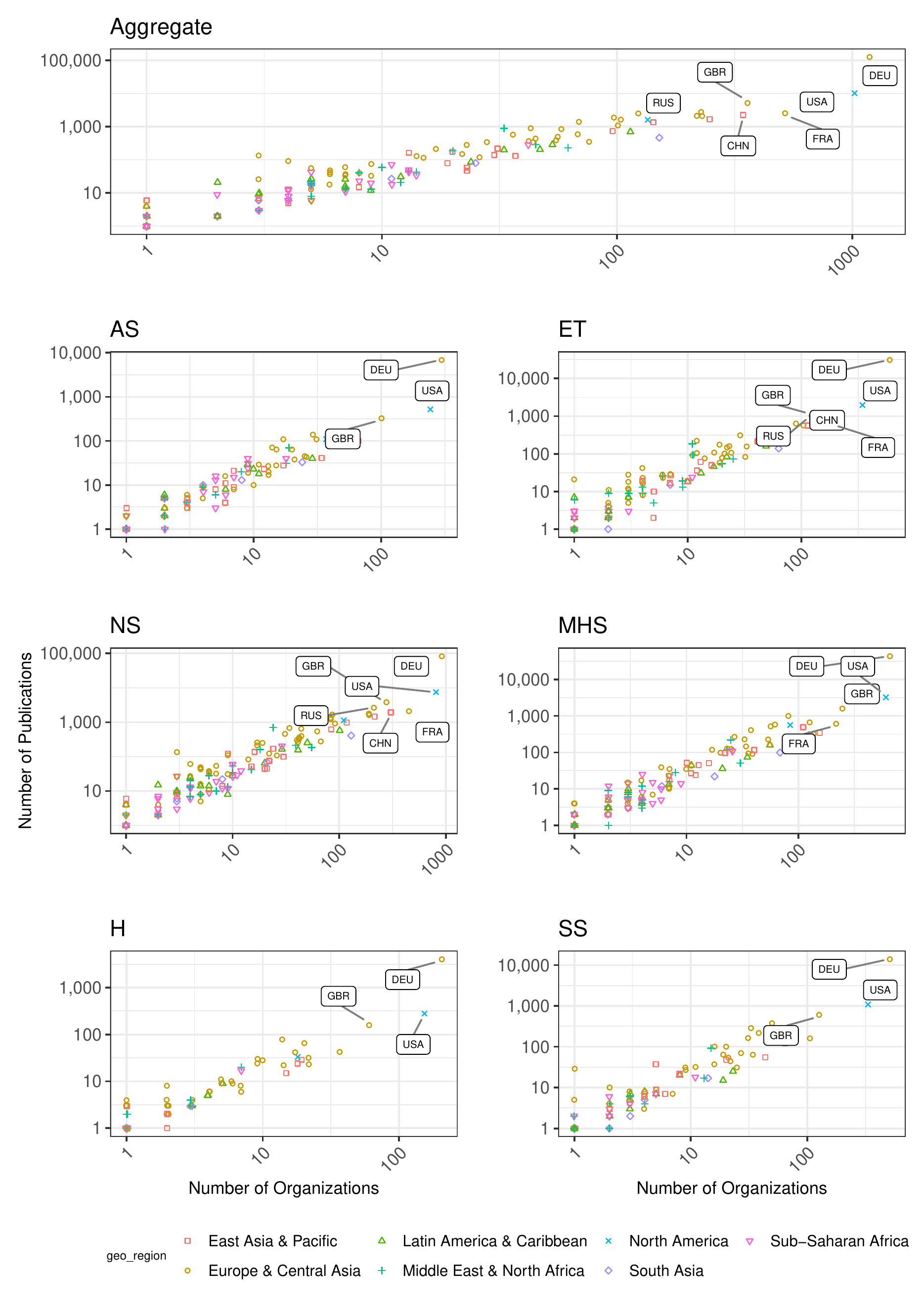} 

}

\caption{Number of organizations and co-authorship with Berlin by countries worldwide, aggregated (top) and by discipline (bottom, 1996-2017, Scopus, X and Y on log scale, label: top 5 percent)}\label{fig:orgs-count-vs-pubs-countries-continents}
\end{figure}

Figure \ref{fig:orgs-count-vs-pubs-countries-continents} presents the internationalization of collaborations between OECD disciplines divided over continental and geographical regions worldwide. It also shows the countries where the top five percent of collaborators in terms of number of organizations and publications are located (see country labels). Note that the scale of X and Y axes are different among panels of figure and they are on log 10 scale. Collaborations inside Germany prevail the aggregate image in all disciplines. After Germany, the USA, the UK, France, China and Russia dominate the aggregate view with highest number of publications and organizations. However, disciplines have noticeable differences. In AS, H and SS, Germany, the USA and the UK are the countries where most of collaborators and prolific ones are located. In MHS, France is the only country which joins the previously mentioned top five percent group. In NS and ET, China and Russia join this group as well and the image becomes closest to the aggregate level. This further improves our findings for \textbf{RQ2} and \textbf{RQ3}.

\begin{landscape}

\begin{figure}

{\centering \includegraphics[width=1\linewidth]{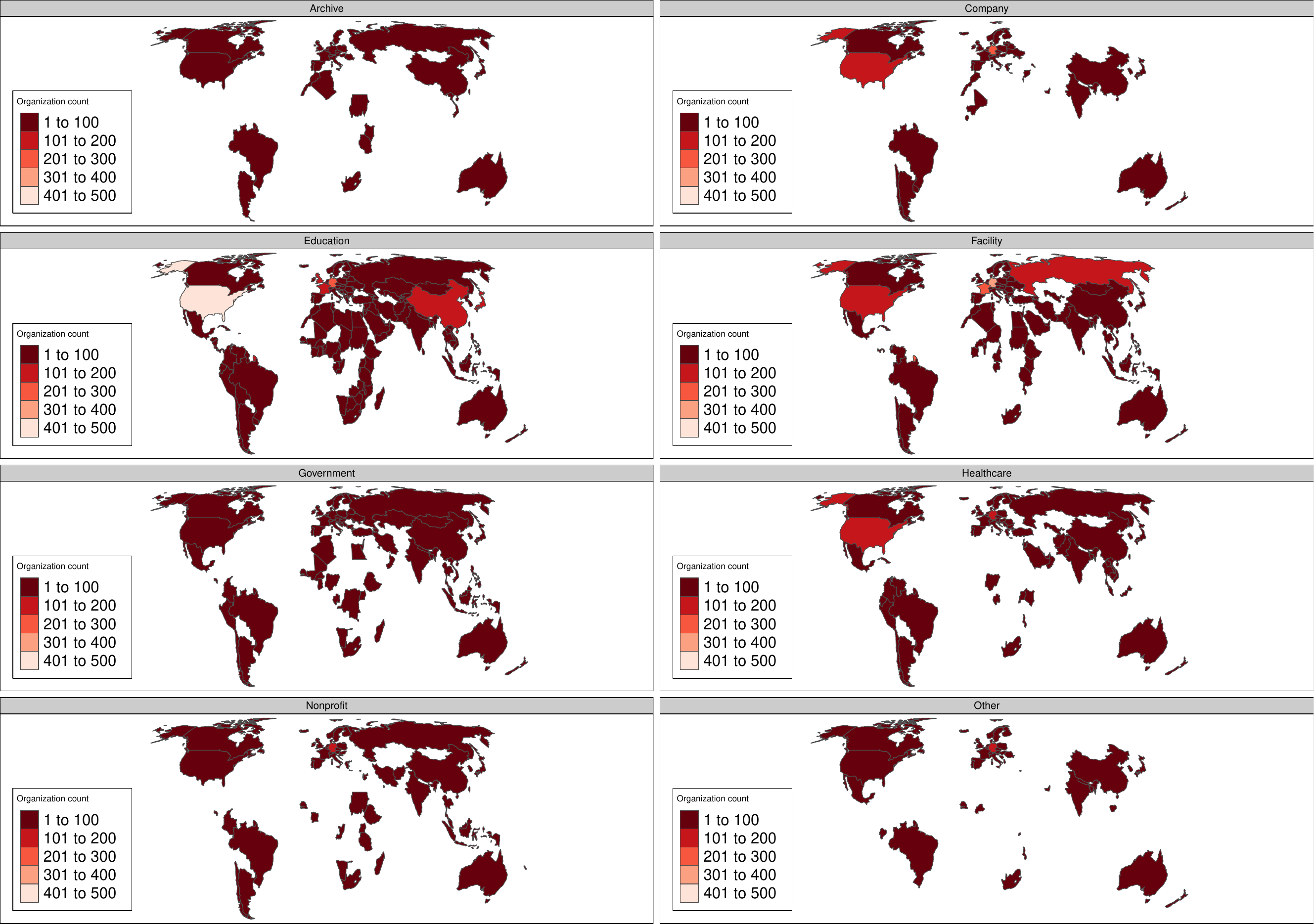} 

}

\caption{Countries worldwide collaborating with Berlin region by sector (color: N. of organizations. If a country does not have presence in a sector, it is removed)}\label{fig:map-of-orgs-type-world}
\end{figure}

\end{landscape}

\begin{landscape}

\begin{figure}

{\centering \includegraphics[width=1\linewidth]{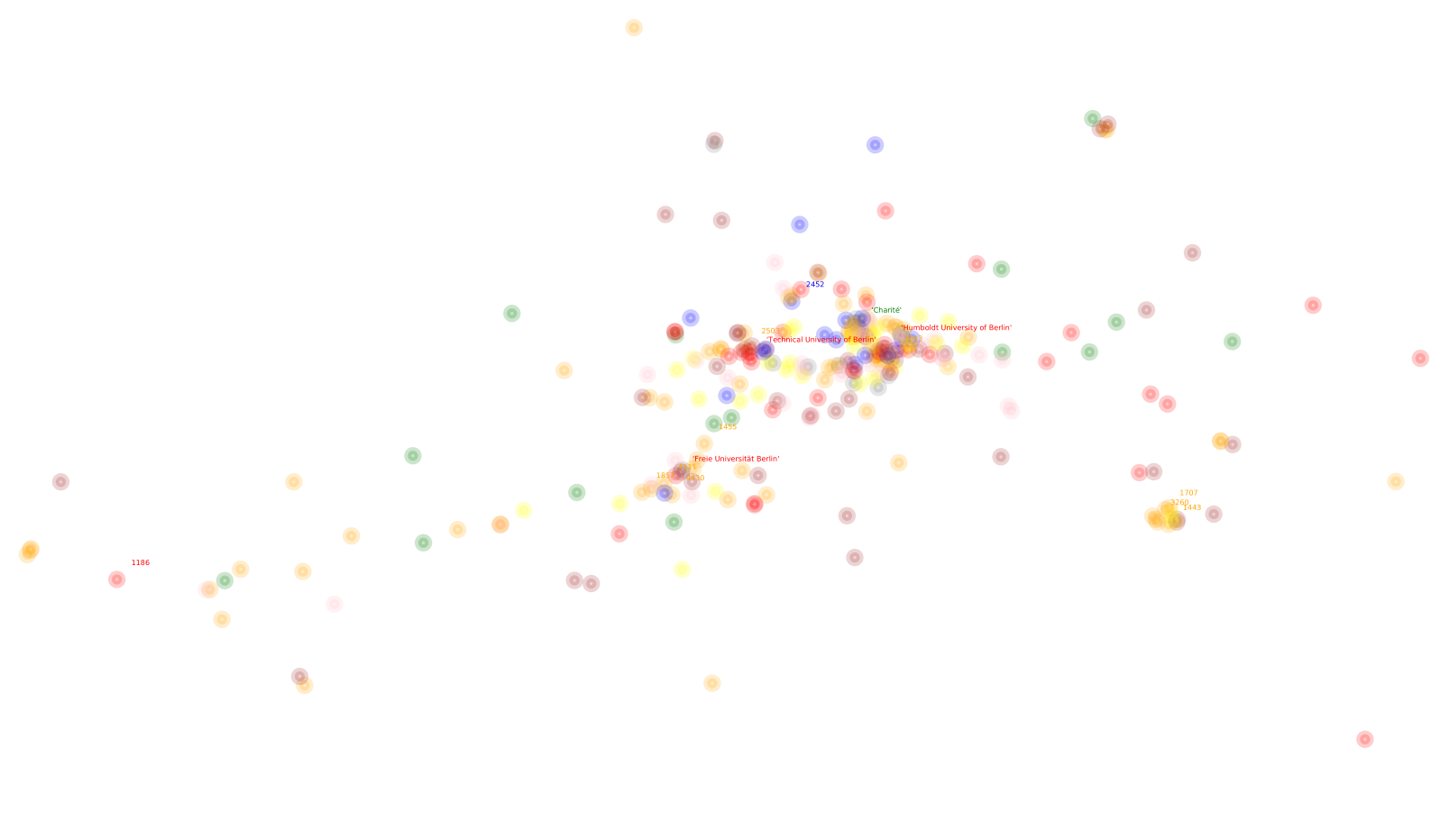} 

}

\caption{Organizations in Berlin (Colors: Education = red, Nonprofit = yellow, Government = blue, Facility = orange, Healthcare = green, Company = brown, Other = pink, Archive = gray, NA = white, Labels: > 10,000 publication, name, 1000-10,000 publication, count)}\label{fig:map-of-orgs-type-germany-berlin}
\end{figure}

\end{landscape}

To investigate our \textbf{RQ4} and \textbf{RQ5}, we focus on organization sectors. In total out of 7,257 unique organizations in Berlin sample based on ROR, there were 2,844 from education sector, 1,667 facility, 860 healthcare, 587 company, 436 nonprofit, 429 government, 282 Other, 124 archive and 28 not available. Table \ref{tab:description-different-org-types} presents the distribution of organizations in different sectors in five countries with highest number of organizations (i.e., China, Germany, France, the UK and the USA, in alphabetical order). While education is the sector with the highest number of organizations in four countries, facility has the highest number of organizations in Germany which can be an artifact of the disambiguation and exclusion of publications with non-disambiguated organizations. Figure \ref{fig:map-of-orgs-type-world} presents the geographical distribution of organizations worldwide separated by sectors and aggregated in countries. To make the image clearer, we remove countries where no organizations from a given sector is present and brighter colors show higher number of organizations in a given country. It is clear that most countries have organizations in education sector. Another evident pattern is that more developed countries (e.g., in western Europe, North America and Oceania) have representatives in all sectors which signals the higher sectoral diversity of the science systems of these countries. However, China and India are two specific cases outside of previously mentioned regions with representation in many sectors. The distribution of companies is another interesting observation where many countries do not have any representatives in contrast to education sector. Figure \ref{fig:map-of-orgs-type-germany-berlin} focuses on Berlin metropolitan region providing a more fine-grained view of the sectoral distribution of organizations and their dense geographical proximity. Note that on this figure, name of organizations with more than 10,000 publications is printed which are the four BUA members. Berlin presents a science hub with high degree of densely located academic and non-academic scientific organizations which belong to multiple sectors and based on previous map in figure \ref{fig:map-of-orgs-type-world}, they collaborate with organizations from many sectors worldwide.

\begin{table}

\caption{\label{tab:description-different-org-types}Five countries with highest number of organizations by sector (GRID data based on Berlin sample 1996-2017)}
\centering
\fontsize{9}{11}\selectfont
\begin{tabular}[t]{l|l|r}
\hline
Country code & Organization sector & Count\\
\hline
CHN & Education & 193\\
\hline
CHN & Facility & 77\\
\hline
CHN & Healthcare & 39\\
\hline
CHN & Government & 18\\
\hline
CHN & Company & 7\\
\hline
CHN & Nonprofit & 5\\
\hline
CHN & Other & 3\\
\hline
CHN & Archive & 1\\
\hline
DEU & Facility & 319\\
\hline
DEU & Education & 215\\
\hline
DEU & Company & 205\\
\hline
DEU & Healthcare & 115\\
\hline
DEU & Nonprofit & 113\\
\hline
DEU & Other & 113\\
\hline
DEU & Government & 66\\
\hline
DEU & Archive & 35\\
\hline
FRA & Facility & 261\\
\hline
FRA & Education & 114\\
\hline
FRA & Healthcare & 45\\
\hline
FRA & Government & 35\\
\hline
FRA & Company & 34\\
\hline
FRA & Other & 15\\
\hline
FRA & Nonprofit & 8\\
\hline
FRA & Archive & 4\\
\hline
GBR & Education & 114\\
\hline
GBR & Healthcare & 91\\
\hline
GBR & Facility & 46\\
\hline
GBR & Company & 34\\
\hline
GBR & Nonprofit & 26\\
\hline
GBR & Government & 25\\
\hline
GBR & Other & 16\\
\hline
GBR & Archive & 7\\
\hline
USA & Education & 423\\
\hline
USA & Healthcare & 157\\
\hline
USA & Company & 129\\
\hline
USA & Facility & 110\\
\hline
USA & Nonprofit & 95\\
\hline
USA & Government & 44\\
\hline
USA & Other & 34\\
\hline
USA & Archive & 23\\
\hline
\end{tabular}
\end{table}

\newpage

\hypertarget{structure-of-institutional-scientific-collaborations-in-berlin}{%
\subsection{Structure of institutional scientific collaborations in Berlin}\label{structure-of-institutional-scientific-collaborations-in-berlin}}

We focus now on communities of co-authorship identified from the giant component using bipartite community detection (\textbf{RQ6} and \textbf{RQ7}). This enables us to go beyond the macro descriptive view presented thus far and investigate the structure of collaborations in single publication level. Note that these are communities detected from the giant component which is connected in itself, however, these communities signal the denser areas of the collaboration network. We are interested to know what could be the underlying factors behind these higher densities which constructs these cohesive sub-groups. Note also that we tested with a diverse array of resolution parameters as discussed in the \protect\hyperlink{datamethods}{Data and methods} section. We finally set the resolution parameters that gave the most consistent number of communities.

\begin{figure}

{\centering \includegraphics[width=1\linewidth,]{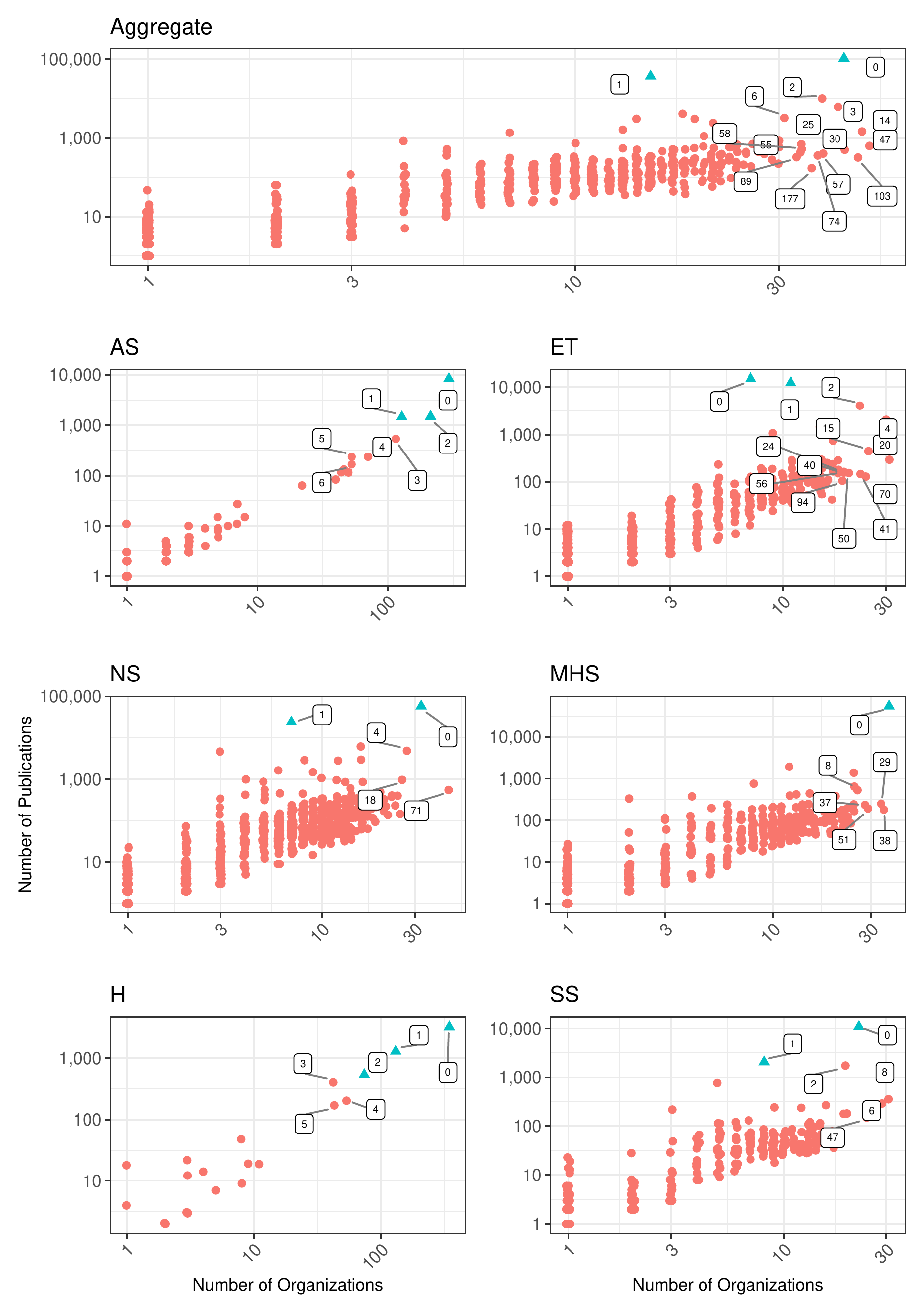} 

}

\caption{Organizations in communities of the giant component vs publication (label: largest and most prolific communities, color and shape: green triangle includes BUA member(s), X and Y on log scale)}\label{fig:composition-of-clusters}
\end{figure}

Figure \ref{fig:composition-of-clusters} presents the distribution of communities based on the number of organizations in each community and aggregate number of publications of all organizations in a given community (each community is represented with one dot). The clearest observation in this figure is the discipline-based collaboration patterns among BUA members indicated by the shape and color of dots where green triangles show presence of one or more BUA member(s) in a given community. In the aggregate view on top, ET, NS and SS, we see BUA members populating the \emph{two} most prolific communities. MHS is the only case where all these four members are present in \emph{a single community} which is due to closer cooperation among BUA members that was formed through a shared faculty by HU and FU located in Charité in 2003 and it seems they have been successful in integrating TU in the collaboration structure. In case of AS and H they populate \emph{three} distinct communities which signals a larger divide in the collaboration structure despite the smaller size and lower number of players in these two disciplines. It is in-line with the self represented research profiles of the BUA members since they aim to have strong research focus in these areas. However, it shows a divide between them that can be bridged by shared research projects or new organizational forms as it was the case in MHS. In ROR, TU is a member of community 1 while FU, CH and HU are members of community 0. In AS, HU and FU are members of community 0 while TU is in community 1 and CH is in community 2. This could be due to the fact that TU, being a technical university, pursues more technical and application oriented research. But the divide between CH from one side and HU and FU on the other side needs further probes for underlying causes. In ET, TU is in community 0 while FU, HU and CH are in community 1. In NS and SS, CH, FU and HU are in community 0 and TU is in community 1. In H, HU and CH are in community 0, FU is in community 1 and TU is in community 2. Overall, it seems that the joint cooperation between HU and FU in form of the shared faculty located in CH is paying off in most cases in form of a more cohesive structure of collaborations. However, in Humanities, where HU, FU and TU have strong research focuses, they have formed distinctive and separated collaboration structures. Of course it can be affected by the disciplinary properties of Humanities which is closer to the ideal type of ``sole investigators'' (Leahey, \protect\hyperlink{ref-leaheySoleInvestigatorTeam2016}{2016}). BUA needs to integrate TU further into the structure of scientific collaborations in the region through shared projects or organizational forms. It is clear from aggregate and disciplinary views that not all communities are populated with the most prolific organizations (in terms of number of publications). There are communities of different sizes consisting of organizations with different levels of productivity (e.g., see the difference between communities 0, 1 and 47 on the aggregate view or communities 0, 1 and 2 in AS). Another clear finding which is in-line with the previously observed number of connected components is that some disciplines present more groupings (e.g., see the distribution of dots in MHS and NS) while some present a lower rate of groupings (e.g., see H and AS). We now exclude communities with the lowest number of organizations which were not prolific and focus instead on the most interesting communities (those with labels in figure \ref{fig:composition-of-clusters}). We present a view of the potential underlying factors based on geographical regions, communities of co-authorship, sectors and productivity which could have lead to the emergence of observed groupings.

\begin{figure}

{\centering \includegraphics[width=1\linewidth,]{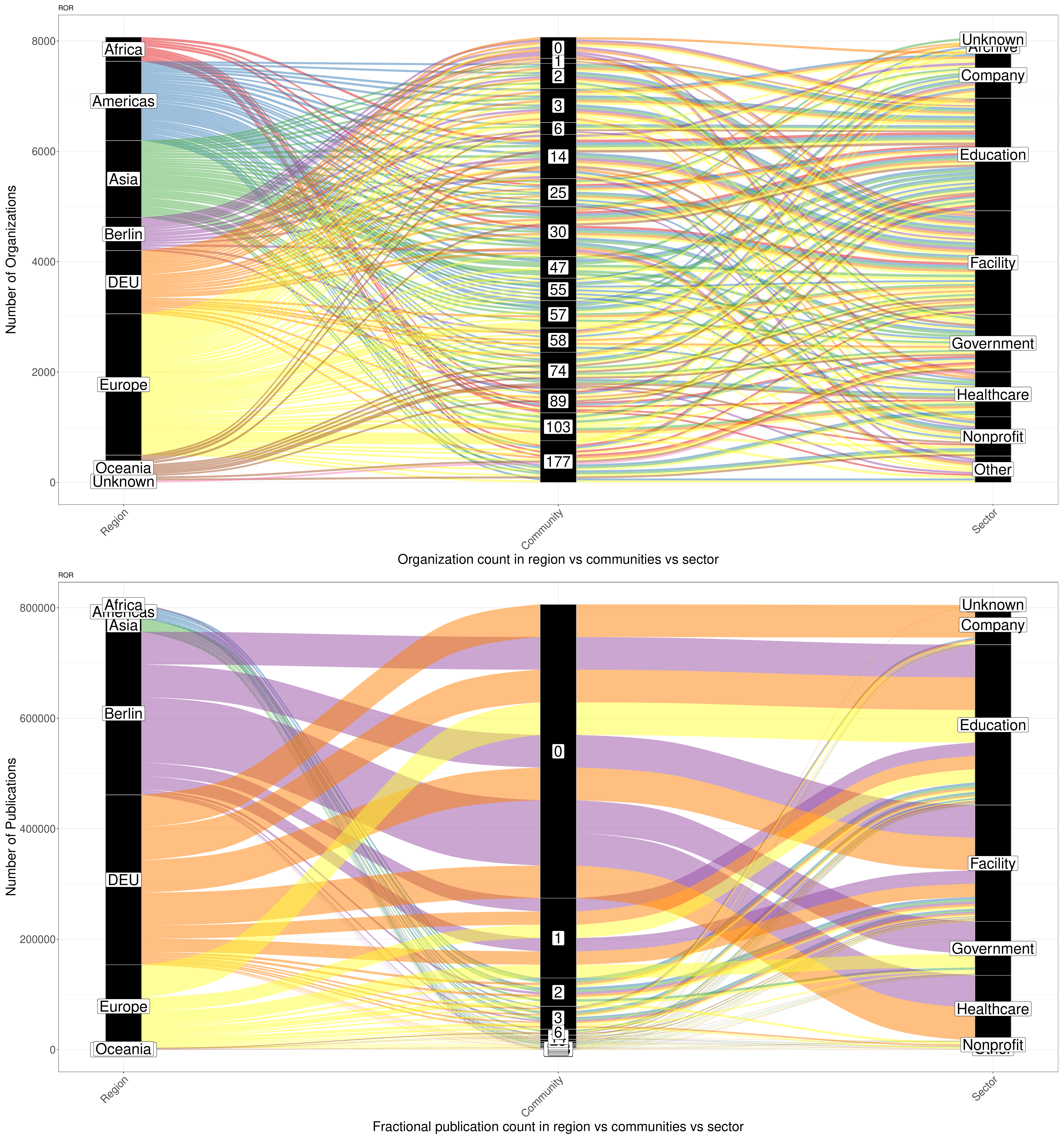} 

}

\caption{Organizations by regions vs membership in communities of the giant component aggregated in ROR (Top: size of communities, Bottom: productivity)}\label{fig:composition-of-clusters-of-continents-ror}
\end{figure}

Figure \ref{fig:composition-of-clusters-of-continents-ror} presents the communities in the aggregate view (i.e., ROR) consisting of more than 30 organizations or with more than 25,000 publications (aggregate of all community members) and Table \ref{tab:composition-of-clusters-of-continents-ror-table} provides details on regional and sectoral composition of these selected communities. We chose these parameters to present the most interesting communities with highest productivity in terms of number of publications and the most distinctive structure of collaborations. Note that, although the most interesting communities are chosen based on number of organizations and \emph{raw count} of publications, but, the total publications presented on the bottom panel of figure \ref{fig:composition-of-clusters-of-continents-ror} are \emph{fractional counts} to better present the share of collaborative works. We separate \emph{Berlin} and \emph{Germany (DEU)} to provide a better comparison of \emph{intra}/\emph{inter}-regional collaborations. Selected communities in ROR are 16 in total (see the labels on central stacked bar, i.e., 0, 1, 2, 3, 6, 14, 25, 30, 47, 55, 57, 58, 74, 89, 103 and 177). From one hand, top and bottom panels of this figure presents the \emph{regional} composition of the communities (see left stacked bars with region names) and \emph{sector} of the organizations in these communities (see right stacked bars with sector names). On the other hand, top panel shows the size of these communities in terms of number of organizations (see the width of ribbons outgoing from each stacked bar) and bottom panel shows the level of productivity in terms of total number of publications by organizations in each community (see the width of ribbons outgoing from each stacked bar).

In the aggregate level in ROR, Europe (excluding Berlin and Germany) has the highest share of organizations (see length of left stacked bar on top panel) and it is followed by Americas, Asia and Germany (excluding Berlin). However, in productivity, Berlin and Germany overtake the other regions. It is clear that communities 0 and 1 are more prolific than other communities (see the length of central stacked bar in bottom panel of the figure \ref{fig:composition-of-clusters-of-continents-ror}), despite their smaller sizes. Although these 16 communities are composed of an international mixture of organizations, the most prolific communities (i.e., 0 and 1) are dominated by Berlin, Germany and other European organizations (compare the multiple ribbons ingoing to each community in central stacked bar on top panel and how the most prolific ones in bottom panel are dominated with few ribbons and highest length of stacked bars). Furthermore, these two communities include all of the four BUA members (FU, CH and HU are in community 0 and TU is in community 1). Organizations from other countries (even those with comparable number of organizations) populate the smaller, less prolific communities. While education and facility have the largest share of organizations and productivity, these communities present a diverse sectoral composition and they are not dominated by any specific sector. Nevertheless, it is interesting that government organizations have representatives in all these selected communities except 6 (see Table \ref{tab:composition-of-clusters-of-continents-ror-table}).

\begin{table}

\caption{\label{tab:composition-of-clusters-of-continents-ror-table}Composition of the largest and most prolific communities of the giant component by region and sector in ROR (N = community size, P = aggregate publications)}
\centering
\resizebox{\linewidth}{!}{
\begin{tabular}[t]{r|r|r|r|r|r|r|r|r|r|r|r|r|r|r|r|r|r|r|r}
\hline
\multicolumn{3}{c|}{ } & \multicolumn{8}{c|}{Region} & \multicolumn{9}{c}{Sector} \\
\cline{4-11} \cline{12-20}
cluster & N & P & Africa & Americas & Asia & Berlin & DEU & Europe & No region & Oceania & Archive & Company & Education & Facility & Government & Healthcare & No sector & Nonprofit & Other\\
\hline
0 & 43 & 104,247 &  &  &  & 6 & 34 & 3 &  &  &  & 1 & 28 & 5 & 1 & 8 &  &  & \\
\hline
1 & 15 & 36,918 &  &  &  & 6 & 6 & 3 &  &  &  &  & 8 & 6 & 1 &  &  &  & \\
\hline
2 & 38 & 9,862 &  & 8 & 6 & 1 & 6 & 17 &  &  &  &  & 24 & 11 & 2 &  &  & 1 & \\
\hline
3 & 41 & 6,074 &  & 5 & 6 & 1 & 7 & 22 &  &  &  & 4 & 20 & 11 & 2 &  &  & 2 & 2\\
\hline
6 & 31 & 3,193 &  & 3 & 2 & 1 &  & 24 &  & 1 &  & 1 & 19 & 11 &  &  &  &  & \\
\hline
14 & 47 & 1,447 & 1 & 6 & 26 & 3 & 3 & 6 &  & 2 &  & 2 & 30 & 8 & 1 & 1 &  & 4 & 1\\
\hline
25 & 34 & 686 & 1 & 13 & 2 & 1 & 5 & 12 &  &  & 1 & 2 & 20 & 7 & 2 & 1 &  & 1 & \\
\hline
30 & 43 & 497 & 4 & 7 & 16 & 1 & 5 & 9 &  & 1 & 1 &  & 19 & 13 & 3 &  & 1 & 3 & 3\\
\hline
47 & 49 & 624 &  & 2 & 43 &  &  & 4 &  &  &  & 7 & 30 & 9 & 1 & 1 &  & 1 & \\
\hline
55 & 34 & 424 &  & 23 & 3 &  &  & 8 &  &  &  & 5 & 5 & 12 & 2 & 5 &  & 5 & \\
\hline
57 & 38 & 398 &  & 5 & 15 &  & 1 & 17 &  &  & 1 & 3 & 16 & 12 & 3 & 2 &  &  & 1\\
\hline
58 & 34 & 519 &  & 4 &  & 1 & 3 & 26 &  &  & 1 & 4 & 8 & 8 & 3 & 9 &  &  & 1\\
\hline
74 & 37 & 356 &  & 3 & 8 & 1 & 1 & 7 &  & 17 &  & 3 & 10 & 7 & 4 & 7 &  & 5 & 1\\
\hline
89 & 33 & 322 & 18 & 3 &  &  & 1 & 10 &  & 1 &  & 2 & 9 & 3 & 4 & 11 &  & 2 & 2\\
\hline
103 & 46 & 317 &  &  & 3 &  & 1 & 42 &  &  &  & 1 & 11 & 15 & 2 & 12 &  & 2 & 3\\
\hline
177 & 36 & 170 & 5 & 6 & 2 &  & 1 & 19 & 1 & 2 & 1 &  & 3 & 4 & 12 & 4 & 1 & 5 & 6\\
\hline
\end{tabular}}
\end{table}

Table \ref{tab:composition-of-clusters-of-continents-AS-table} presents the communities in Agricultural Sciences consisting of more than 50 organizations, which are 7 communities in total (i.e., 0, 1, 2, 3, 4, 5 and 6. These communities can be noted on the sub-panel of figure \ref{fig:composition-of-clusters} dedicated to AS). Similar to the composition observed in aggregate level (i.e., ROR), these communities are composed of an internationalized mixture of organizations. But, different from what we observed in aggregate level (i.e., domination of Berlin, Germany and Europe in the most prolific communities), here we observe that highly prolific communities are composed of a mixture of multiple regions. Considering these internationalized collaborations and the fact that BUA members are located in communities 0 (HU and FU), 1 (TU) and 2 (CH), we can conclude that each of these highly prolific organizations has their own specific group of international collaborators in AS. This is in line with the self-representations of BUA members of their research profiles. Community 0 is the largest and most prolific one in AS. While communities 1 and 2 have comparable productivity rates, the size of community 2 is about twice that of community 1. We cannot observe a dominating pattern of sectors in composition or productivity of the communities in AS and education, facility and government present relatively similar productivity levels which could be an attribute of AS. It is interesting to see the high density of health-care organizations in community 2 where Charité is located.

\begin{table}

\caption{\label{tab:composition-of-clusters-of-continents-AS-table}Composition of the largest and most prolific communities of the giant component by region and sector in AS (N = community size, P = aggregate publications)}
\centering
\resizebox{\linewidth}{!}{
\begin{tabular}[t]{r|r|r|r|r|r|r|r|r|r|r|r|r|r|r|r|r|r|r|r}
\hline
\multicolumn{3}{c|}{ } & \multicolumn{8}{c|}{Region} & \multicolumn{9}{c}{Sector} \\
\cline{4-11} \cline{12-20}
cluster & N & P & Africa & Americas & Asia & Berlin & DEU & Europe & No region & Oceania & Archive & Company & Education & Facility & Government & Healthcare & No sector & Nonprofit & Other\\
\hline
0 & 296 & 8,386 & 32 & 68 & 22 & 12 & 78 & 76 &  & 8 & 20 & 3 & 187 & 49 & 18 & 7 & 1 & 10 & 1\\
\hline
1 & 127 & 1,464 & 9 & 18 & 28 & 11 & 27 & 30 & 2 & 2 &  & 5 & 81 & 26 & 6 & 2 &  & 2 & 5\\
\hline
2 & 210 & 1,498 & 6 & 54 & 35 & 14 & 31 & 65 &  & 5 &  & 10 & 99 & 41 & 13 & 39 & 1 & 7 & \\
\hline
3 & 114 & 538 & 3 & 22 & 12 & 5 & 11 & 56 &  & 5 & 2 & 5 & 51 & 32 & 6 & 5 &  & 9 & 4\\
\hline
4 & 71 & 238 &  & 6 & 9 & 4 & 3 & 42 &  & 7 & 2 & 2 & 37 & 13 & 7 & 4 &  & 3 & 3\\
\hline
5 & 53 & 237 &  & 13 & 13 & 3 & 6 & 18 &  &  & 5 &  & 29 & 13 & 2 &  & 1 & 2 & 1\\
\hline
6 & 53 & 168 & 4 & 8 & 2 & 2 & 5 & 32 &  &  &  & 2 & 23 & 12 & 9 & 3 &  & 2 & 2\\
\hline
\end{tabular}}
\end{table}

Table \ref{tab:composition-of-clusters-of-continents-ET-table} presents the communities in Engineering Technology consisting of more than 18 organizations or those with more than 10,000 publications. These are 13 communities in total (i.e., 0, 1, 2, 4, 15, 20, 24, 40, 41, 50, 56, 70 and 94). On the one hand, Europe, Americas, Asia and Germany seem to have the highest share of organizations populating these communities. On the other hand, education, facility and companies dominate the sectoral composition. The most prolific communities (i.e, 0 and 1 both with more than 12,000 publications) are dominated by Berlin and German organizations and community 1 has only 1 member from Europe. Community 0 includes TU from BUA members. It can signal the specialization of TU in Engineering Technology discipline and a locally oriented structure of collaborations which is in line with what Hoekman et al. (\protect\hyperlink{ref-hoekmanResearchCollaborationDistance2010}{2010}) reported about ET in other European countries. Community 1 composed of 11 organizations from Berlin, Germany and Europe is highly prolific (+12,000 publications) and includes FU, HU and CH from BUA members which signals the closer collaboration ties between these three members. However, community 2 is an interesting case here. It is rather small in size (23 organizations), it has no BUA members and is composed of an international group of organizations and is relatively prolific (the third most prolific with close to +4,000 publications).

\begin{table}

\caption{\label{tab:composition-of-clusters-of-continents-ET-table}Composition of the largest and most prolific communities of the giant component by region and sector in ET (N = community size, P = aggregate publications)}
\centering
\resizebox{\linewidth}{!}{
\begin{tabular}[t]{r|r|r|r|r|r|r|r|r|r|r|r|r|r|r|r|r|r}
\hline
\multicolumn{3}{c|}{ } & \multicolumn{8}{c|}{Region} & \multicolumn{7}{c}{Sector} \\
\cline{4-11} \cline{12-18}
cluster & N & P & Africa & Americas & Asia & Berlin & DEU & Europe & No region & Oceania & Company & Education & Facility & Government & Healthcare & Nonprofit & Other\\
\hline
0 & 7 & 15,116 &  &  &  & 5 & 2 &  &  &  &  & 2 & 5 &  &  &  & \\
\hline
1 & 11 & 12,457 &  &  &  & 4 & 6 & 1 &  &  &  & 5 & 2 & 1 & 3 &  & \\
\hline
2 & 23 & 4,092 &  & 2 & 6 & 1 & 4 & 9 & 1 &  &  & 9 & 9 & 3 & 1 & 1 & \\
\hline
4 & 30 & 2,059 &  & 1 & 7 & 1 & 2 & 19 &  &  & 1 & 18 & 6 & 2 &  & 1 & 2\\
\hline
15 & 25 & 446 &  & 2 & 3 &  & 1 & 19 &  &  & 1 & 10 & 11 & 1 & 1 & 1 & \\
\hline
20 & 31 & 294 &  & 2 & 2 &  & 3 & 23 &  & 1 & 2 & 13 & 7 & 1 & 2 & 4 & 2\\
\hline
24 & 19 & 163 & 1 & 5 & 1 &  & 1 & 11 &  &  & 2 & 10 & 7 &  &  &  & \\
\hline
40 & 19 & 158 &  &  & 11 &  & 1 & 7 &  &  & 5 & 8 & 6 &  &  &  & \\
\hline
41 & 23 & 147 &  & 13 & 1 & 2 & 2 & 5 &  &  & 2 & 10 & 5 & 1 & 2 & 3 & \\
\hline
50 & 20 & 154 &  &  & 16 &  &  & 4 &  &  & 2 & 15 & 2 & 1 &  &  & \\
\hline
56 & 19 & 158 &  & 2 & 3 &  &  & 13 & 1 &  & 1 & 8 & 6 & 1 & 1 & 2 & \\
\hline
70 & 24 & 129 &  & 4 & 5 &  &  & 14 &  & 1 & 3 & 14 & 4 & 3 &  &  & \\
\hline
94 & 19 & 106 &  & 12 & 2 &  & 1 & 4 &  &  &  & 15 & 2 &  &  & 1 & 1\\
\hline
\end{tabular}}
\end{table}

Table \ref{tab:composition-of-clusters-of-continents-NS-table} presents the communities in Natural Sciences consisting of more than 25 organizations or those with more than 20,000 publications. These criteria are satisfied only by five communities (i.e., 0, 1, 4, 18 and 71). Community 0 is the most prolific one (58,586 publications) and \emph{only} includes 32 members which are all from Berlin and Germany with one member from Europe. The second most prolific community with more than 24,000 publications is community 1 with seven members which are all from Berlin and Germany with one organization from Europe. Community 71 is the largest one with 45 members from Europe (education, facility, government, health-care, non-profit and other), Americas (education, facility and health-care sectors), Asia (education and facility), Africa (facility) and Oceania (facility) \emph{without} any members from Berlin or Germany. Note that, these communities are detected based on denser areas of the giant component and they do not necessarily need to include Berlin organizations. However, while being the largest community, it is not prolific compared to other NS communities (553 publications). Community 18 is another interesting case which is highly international with 26 members and 968 publications. Africa is only present in community 71 and Oceania is only present in communities 18 and 71 i.e., the communities with the lowest productivity levels. NS is showing a highly prolific community (i.e., 0, 32 members, 3 BUA members, i.e., HU, FU and CH) and a small but still prolific community (i.e., 1, 7 members, includes TU from BUA members) with BUA members. This can signal a divide between these two groups of BUA members which are collaborating within Berlin, Germany and Europe but with less overlapping collaboration ties that can be bridged and fostered through future cooperations.

\begin{table}

\caption{\label{tab:composition-of-clusters-of-continents-NS-table}Composition of the largest and most prolific communities of the giant component by region and sector in NS (N = community size, P = aggregate publications)}
\centering
\resizebox{\linewidth}{!}{
\begin{tabular}[t]{r|r|r|r|r|r|r|r|r|r|r|r|r|r|r|r|r}
\hline
\multicolumn{3}{c|}{ } & \multicolumn{7}{c|}{Region} & \multicolumn{7}{c}{Sector} \\
\cline{4-10} \cline{11-17}
cluster & N & P & Africa & Americas & Asia & Berlin & DEU & Europe & Oceania & Company & Education & Facility & Government & Healthcare & Nonprofit & Other\\
\hline
0 & 32 & 58,586 &  &  &  & 6 & 25 & 1 &  & 1 & 20 & 4 & 1 & 6 &  & \\
\hline
1 & 7 & 24,045 &  &  &  & 4 & 2 & 1 &  &  & 2 & 4 & 1 &  &  & \\
\hline
4 & 27 & 4,869 &  & 2 & 6 & 1 & 2 & 16 &  & 2 & 17 & 6 & 1 &  &  & 1\\
\hline
18 & 26 & 968 &  & 2 & 16 & 3 & 1 & 3 & 1 &  & 20 & 1 & 1 &  & 2 & 2\\
\hline
71 & 45 & 553 & 1 & 3 & 2 &  &  & 38 & 1 &  & 9 & 19 & 3 & 6 & 4 & 4\\
\hline
\end{tabular}}
\end{table}

Table \ref{tab:composition-of-clusters-of-continents-MHS-table} presents the communities in Medical and Health Sciences with more than 25 organizations. These are six communities (i.e., 0, 8, 29, 37, 38 and 51). MHS is presenting a highly interesting case where community 0 is the most prolific (55,480 publications) and it is populated by only 37 organizations from Berlin and Germany (education, facility, health-care, in both Germany and Berlin. One government organization from Berlin and one company from Germany i.e., excluding Berlin). Community 0 includes all four BUA members and signals a high level of intraregional collaboration in MHS. This shows that the strategic cooperation between FU and HU to establish a shared MHS faculty in Charité has paid off and they have been successful in integrating TU and three other organizations from Berlin and 30 organizations from Germany. All other communities, whether international or regional have much lower productivity levels compared to community 0.

\begin{table}

\caption{\label{tab:composition-of-clusters-of-continents-MHS-table}Composition of the largest and most prolific communities of the giant component by region and sector in MHS (N = community size, P = aggregate publications)}
\centering
\resizebox{\linewidth}{!}{
\begin{tabular}[t]{r|r|r|r|r|r|r|r|r|r|r|r|r|r|r|r|r|r}
\hline
\multicolumn{3}{c|}{ } & \multicolumn{7}{c|}{Region} & \multicolumn{8}{c}{Sector} \\
\cline{4-10} \cline{11-18}
cluster & N & P & Africa & Americas & Asia & Berlin & DEU & Europe & No region & Archive & Company & Education & Facility & Government & Healthcare & Nonprofit & Other\\
\hline
0 & 37 & 55,480 &  &  &  & 7 & 30 &  &  &  & 1 & 23 & 5 & 1 & 7 &  & \\
\hline
8 & 26 & 529 & 2 & 18 & 2 &  &  & 4 &  &  & 2 & 7 & 4 &  & 12 &  & 1\\
\hline
29 & 34 & 251 &  & 1 &  &  & 1 & 32 &  &  &  & 12 & 9 & 2 & 9 & 2 & \\
\hline
37 & 28 & 234 &  & 4 & 1 &  & 1 & 21 & 1 &  &  & 12 & 1 & 2 & 11 & 1 & 1\\
\hline
38 & 35 & 180 &  & 5 & 28 &  &  & 2 &  &  & 3 & 24 & 5 & 1 & 2 &  & \\
\hline
51 & 29 & 192 & 16 & 4 &  &  &  & 9 &  & 1 & 2 & 10 & 2 & 1 & 8 & 2 & 3\\
\hline
\end{tabular}}
\end{table}

Table \ref{tab:composition-of-clusters-of-continents-H-table} presents the communities in Humanities consisting of more than 30 organizations. These are six communities (i.e., 0, 1, 2, 3, 4 and 5). Community 0 is composed of a highly international group of organizations from all regions. Other communities in Humanities are comparably international, however they are less prolific than community 0. Similar to our expectation, Humanities scholars are mainly affiliated to organizations in the education sector, but, a highly inter-sectoral composition is observed as well. BUA members are located in \emph{three} most prolific communities i.e., 0 (two BUA members, HU and CH), 1 (FU) and 2 (TU) which signals a rather non-overlapping structure of collaboration which is consisting of highly internationalized mixture of organizations. It is in line with high focus of BUA members on distinctive areas in H mentioned in their self-representations in BUA proposed (Berlin University Alliance, \protect\hyperlink{ref-berlinuniversityallianceGemeinsamImVerbund2018}{2018}, \protect\hyperlink{ref-berlinuniversityallianceBerlinUniversityAlliance2019}{2019}) and can be remnants of the era where BUA members needed to have mutually exclusive areas of focus and identities. Note that, as presented in figure \ref{fig:single-multiple-country-publications-disciplines}, Humanities is not that much internationalized and more than 75\% of all publications are \emph{single country} ones (i.e., intra-Germany co-authorship). Nevertheless, Humanities so far is presenting the closest image to the one observed in Agricultural Sciences where Berlin, Germany and European organizations do not dominate the prolific communities and we observe an internationalized mixture or organizations from multiple sectors.

\begin{table}

\caption{\label{tab:composition-of-clusters-of-continents-H-table}Composition of the largest and most prolific communities of the giant component by region and sector in H (N = community size, P = aggregate publications)}
\centering
\resizebox{\linewidth}{!}{
\begin{tabular}[t]{r|r|r|r|r|r|r|r|r|r|r|r|r|r|r|r|r|r|r}
\hline
\multicolumn{3}{c|}{ } & \multicolumn{8}{c|}{Region} & \multicolumn{8}{c}{Sector} \\
\cline{4-11} \cline{12-19}
cluster & N & P & Africa & Americas & Asia & Berlin & DEU & Europe & No region & Oceania & Archive & Company & Education & Facility & Government & Healthcare & Nonprofit & Other\\
\hline
0 & 343 & 3,319 & 11 & 92 & 23 & 20 & 60 & 120 & 3 & 14 & 1 & 4 & 260 & 43 & 13 & 14 & 4 & 4\\
\hline
1 & 130 & 1,294 & 2 & 28 & 20 & 5 & 24 & 46 & 1 & 4 & 1 & 2 & 85 & 22 & 12 & 4 & 3 & 1\\
\hline
2 & 74 & 549 & 1 & 31 & 5 & 4 & 12 & 20 &  & 1 & 2 & 1 & 52 & 9 & 1 & 2 & 5 & 2\\
\hline
3 & 42 & 411 & 1 & 17 & 1 & 6 & 1 & 16 &  &  & 2 & 1 & 25 & 5 & 1 &  & 6 & 2\\
\hline
4 & 53 & 206 &  & 3 & 11 & 3 & 4 & 32 &  &  & 3 &  & 26 & 20 & 4 &  &  & \\
\hline
5 & 43 & 169 &  & 4 & 2 & 4 & 10 & 22 & 1 &  & 5 & 1 & 19 & 10 & 1 & 2 & 4 & 1\\
\hline
\end{tabular}}
\end{table}

Table \ref{tab:composition-of-clusters-of-continents-SS-table} presents the communities in Social Sciences consisting of more than 20 organizations or those with more than 1,500 publications. These are six communities in total (i.e., 0, 1, 2, 6, 8 and 47). Social Sciences are presenting an image far from the one observed in the case of Humanities and closer to natural and hard sciences. The two most prolific communities (i.e., 0 and 1) are completely dominated by Berlin, German and European organizations. Three of the BUA members are located in community 0 (FU, HU and CH) and one in community 1 (TU). Since FU and HU have formed a closer collaboration with CH since 2003, it would be interesting to look further into the topics of focus in their research to investigate whether these separate communities which are all active in SS study different subjects. Both these communities are collaborating preferably within Berlin and Germany and they have only three members from Europe. Community 0 with 22 members is the most prolific (10,895 publications). Community 2 is an interesting case with one organization from Berlin (which is not from BUA members), six organizations from Americas, four from Germany, seven from Europe and in relative terms, it has a high productivity. It can signal even higher disciplinary divide in Humanities in the Berlin region. However, there are smaller, more international communities which are not highly prolific as it was the case in ET, NS and MHS. Americas have the largest share of organizations after Europe and it is highly represented in community 8 which is only composed of Americas, two organizations from Berlin and 11 organizations from Europe (excluding Germany).

\begin{table}

\caption{\label{tab:composition-of-clusters-of-continents-SS-table}Composition of the largest and most prolific communities of the giant component by region and sector in SS (N = community size, P = aggregate publications)}
\centering
\resizebox{\linewidth}{!}{
\begin{tabular}[t]{r|r|r|r|r|r|r|r|r|r|r|r|r|r|r|r|r}
\hline
\multicolumn{3}{c|}{ } & \multicolumn{7}{c|}{Region} & \multicolumn{7}{c}{Sector} \\
\cline{4-10} \cline{11-17}
cluster & N & P & Africa & Americas & Asia & Berlin & DEU & Europe & No region & Company & Education & Facility & Government & Healthcare & Nonprofit & Other\\
\hline
0 & 22 & 10,895 &  &  &  & 3 & 17 & 2 &  & 1 & 14 & 2 &  & 5 &  & \\
\hline
1 & 8 & 2,046 &  &  &  & 4 & 3 & 1 &  &  & 3 & 4 &  &  &  & 1\\
\hline
2 & 19 & 1,733 &  & 6 &  & 1 & 4 & 7 & 1 &  & 17 & 2 &  &  &  & \\
\hline
6 & 29 & 291 & 1 & 4 & 4 & 1 & 1 & 18 &  & 1 & 16 & 9 & 1 & 1 &  & 1\\
\hline
8 & 31 & 353 &  & 18 &  & 2 &  & 11 &  & 2 & 24 & 2 &  &  & 2 & 1\\
\hline
47 & 24 & 150 & 1 & 3 & 2 &  & 2 & 16 &  & 1 & 16 & 2 & 4 & 1 &  & \\
\hline
\end{tabular}}
\end{table}

\hypertarget{conclusions}{%
\section{Discussion}\label{conclusions}}

In this paper, we provide a quantitative, exploratory and macro view of the structure of scientific collaborations in the Berlin metropolitan region. Our main level of analysis was scientific \emph{organizations} (which can be academic or non-academic organizations or firms) and we investigated the share of collaborative work, internationalized work versus single country collaborations. We covered all OECD scientific disciplines and presented a comparative view of the similarities and differences of collaborations in these disciplines.

In methodological terms, we developed two organization name disambiguation techniques (i.e., PyString and Fuzzy matching) and compared their performance and coverage with an established technique (i.e., Research Organization Registry (ROR)). We presented the high impact organization name disambiguation could have on the constructed collaborations networks and how it can bias measures and trends. We had to exclude 51\% of the publications which had one or more non-disambiguated organizations to limit our analysis to successfully disambiguated cases.

At a first view and only based on descriptive analysis, we observed a highly collaborative scientific landscape. Some disciplines present a high degree of difference between \emph{raw} and \emph{fractional} count of publications. Despite the prevalence of collaborative works and increasing trend towards internationalization in aggregate view, we observed that some disciplines (e.g., Natural Sciences and Agricultural Sciences) have more internationalized collaborations while other disciplines (e.g., Medical and Health Sciences, Humanities and Social Sciences) are less internationalized or they did not present a steady upward trend which is in-line with observation by Moed et al. (\protect\hyperlink{ref-moedInternationalScientificCooperation1991}{1991}) and Babchuk et al. (\protect\hyperlink{ref-babchukCollaborationSociologyOther1999}{1999}). But our further investigation showed that these disciplines have a more complex structure of collaboration which in some cases is highly dominated by local (e.g., Berlin and Germany) organizations and in some cases it is already Europeanized. In rare cases (e.g., Agricultural Sciences and Humanities) we observed a higher representation of international organizations among the most prolific communities despite the smaller size of these disciplines and in case of Humanities less than 25\% share of internationalization. Overall we observed a high degree of collaboration between Berlin organizations and other European countries which is in-line with the trend observed by Hoekman et al. (\protect\hyperlink{ref-hoekmanResearchCollaborationDistance2010}{2010}), however, we observed that the image is more complicated once we look at cohesive subgroups and groupings.

In geographical terms, different disciplines present various collaboration trends, however, all of them have high degrees of regional and European collaboration. Berlin presents a specific case. It is similar to a science hub with a diverse sectoral composition of organizations which is in line with Balland et al. (\protect\hyperlink{ref-ballandComplexEconomicActivities2020}{2020})'s observation in metropolitan regions in the USA and Rammer et al. (\protect\hyperlink{ref-rammerKnowledgeProximityFirm2020}{2020})'s observation of Berlin metropolitan region. However, it can be due to our data gathering strategy where only publications with at least one organization located in Berlin are included. Thus there could be other collaborations between the partners excluding Berlin organizations that we do not cover here. North America is the preferred collaboration partner outside of Europe. But this image changes in some disciplines with closer share of collaborators from East Asia and Pacific and North America and their shares is the closest in Engineering Technology. Intra Germany (i.e., single country) collaborations prevail the co-authorships. The USA and the UK are central members of the collaboration landscape in most disciplines based on the aggregate number of publications. But, France, China and Russia join the top five percent of most prolific collaborators in Engineering Technology and Natural Sciences.

We provided a sectoral view of the geographical distribution of organizations worldwide and in Berlin. Some countries present a highly diverse science system consisting of a wide range of sectors among those collaborating with Berlin organizations. However, in most countries, \emph{education} is the prevailing sector where scientific publications are produced which is not counterintuitive. China and India have representatives in many sectors.

We modeled the scientific collaborations through \emph{bipartite} co-authorship networks which treats each scientific publication as an event where organizations interact in producing scientific texts. Our bipartite community detection configuration was helpful in detecting the diverse composition of organizational teams contributing to scientific publications which could be overlooked if the network is projected to one-mode due to artificially high cliquish behavior. We extracted the largest connected components in the aggregate level and for each scientific discipline. We then looked at denser collaborating groups. We observed that in most disciplines, with the exception of \emph{Humanities} and \emph{Agricultural Sciences}, the most prolific communities were composed of organizations located in the Berlin metropolitan region and they were collaborating either within Berlin or with other German organizations or exclusively with European organizations. There were of course internationalized communities in all disciplines, but they were not highly prolific.

Disciplines present interesting differences in the composition of communities (i.e., denser collaborating groups). Not all communities were composed of highly prolific organizations. However, in some cases we observed that all highly prolific organizations were member of one or two communities. We observed a highly regional oriented Medical and Health Sciences, national and Europe oriented Natural Sciences, Engineering Technology and Social Sciences and an internationally oriented Agricultural Sciences and Humanities. This can be due to the restrictive criteria we set in the further investigation of the highly prolific or the largest communities or it can signal strategic regional coalitions.

Looking at the members of Berlin University Alliance (BUA) and their position in these cohesive sub-groups presented interesting findings. In Agricultural Sciences and Humanities which were the disciplines with lower aggregate internationalization (less than 25\%), we observed higher rates of internationalized collaboration once the structure of communities were analyzed and there was a larger divide between BUA members. They formed denser collaboration ties within \emph{three} different communities. This is in line with the self-representations of the research profiles of the BUA members (Berlin University Alliance, \protect\hyperlink{ref-berlinuniversityallianceGemeinsamImVerbund2018}{2018}, \protect\hyperlink{ref-berlinuniversityallianceBerlinUniversityAlliance2019}{2019}). Only in Medical and Health Sciences which was dominated by high productivity of Berlin and Germany, the four BUA members were located in \emph{one} community and collaborated densely. This is highly affected by the fact that from 2003, HU and FU have jointly established a MHS faculty located in Charité and our findings show that this strategic coalition has been successful in integrating other organizations from Berlin and TU. Furthermore, some of the observed disciplinary divide between BUA members could be remnants of the east-west division in Germany and the reorganization of research profiles and mutually exclusive definition of areas of focus to reduce parallel work and competition that happened after the reunification. This divide was recently observed in biotechnology field by Abbasiharofteh \& Broekel (\protect\hyperlink{ref-abbasiharoftehStillShadowWall2020}{2020}). TU presents a specific case. While the other three BUA members formed different collaboration structures based on disciplinary specialization, TU is in most cases member of a separated community of its own. In ET, TU's collaboration network is dominated by other Berlin and German organizations. It shows that BUA needs to develop further strategic cooperations among the members to ensure a higher integration, similar to the case of MHS. However, this might be due to the fact that we included \emph{conference proceedings} which is a specific publication type preferred more by the technical universities. Since TU is the main technical university in our sample, the collaboration structure reflected in this document type might have affected the observed results and over inflate the divide between TU and other BUA members. Charité from other hand seems to have built a large collaboration network worldwide which is evident in health-care sector and high representation of national and international organizations in the communities where Charité was located. BUA can plan to benchmark this as a case of successful internationalization of collaborations.

We conclude that mixing a macro and global view while keeping regional, national and continental granularity can help in describing observed quantitative trends. It is necessary to move beyond the macro descriptive view presented based on yearly count of publications or increasing trends of team science. In addition, investigation of self-representation of research profiles of scientific organizations was helpful in interpretation of observed trends. As our investigation proved, not all members of the community are moving towards internationalization and some parts of the community, which are normally the highly prolific ones, prevail and distort the aggregate images. This methodological approach is close to what Stadtfeld (\protect\hyperlink{ref-stadtfeldMicroMacroLinkSocial2018}{2018}) suggests in moving between micro, meso and macro levels to better explain the observed trends.

Our paper suffers from certain limitations. When we construct the co-authorship network at the organization level, we naturally overlook the changes that happen in the composition of researchers affiliated to those organizations. The same organization could have a highly different composition of members over the time that affects the type of research carried out and collaboration ties formed.

We only use Scopus as the main database and although it covers German speaking publications, it is dominated by English speaking records. Bibliometric databases, including Scopus, are regularly updated which can affect the temporal trends we observe here. In addition, each bibliometric database covers a specific set of scientific publications (see Stahlschmidt et al. (\protect\hyperlink{ref-stahlschmidtPerformanceStructuresGerman2019}{2019}) for a comparison between WOS and Scopus), despite the similarities, there are differences in philosophies and approaches to what should be indexed. Furthermore, we were unable to disambiguate all the organizations in our sample which lead to excluding 51\% of the publications which had one or more non-disambiguated organizations. Thus, while our results seem to be following the general trends observed in the German scientific system (see Stahlschmidt et al. (\protect\hyperlink{ref-stahlschmidtPerformanceStructuresGerman2019}{2019}), Stephen et al. (\protect\hyperlink{ref-stephenPerformanceStructuresGerman2020}{2020}) and Aman (\protect\hyperlink{ref-amanHowCollaborationImpacts2016}{2016})), but the specificities observed in the structure of scientific collaborations among BUA members and the international collaborations could be highly affected if we were able to improve the coverage of the disambiguation techniques.

Another limitation of our data and research in organization level is the \emph{superstar} researchers with multiple affiliations. We assume that these researchers have received resources from each of these multiple organizations. Thus, we consider these researchers as bridges between these organizations. While in the networks constructed, these cases might seem as an international collaboration when a single author is affiliated to multiple countries. High quality data with disambiguated records of publications in author level would allow a more complete investigation. Another limitation of our study could be that our disambiguation techniques penalizes non-English speaking countries or less known organizations which are usually less prolific. They can be more represented among the organizations that we excluded from our analysis since the disambiguation did not give reliable results for them. In addition, different disambiguation techniques are effective to identify a differing set of organizations and any choice would have implications on a subset of the organizations while penalizing another subset.

We do not have any insight over the background of individual researchers affiliated to these scientific organizations. We do not know about the motivations that drive the scientific collaboration and observed trends (Subramanyam, \protect\hyperlink{ref-subramanyamBibliometricStudiesResearch1983}{1983}; Katz \& Martin, \protect\hyperlink{ref-katzWhatResearchCollaboration1997}{1997}). As an example, in case of all disciplines, we observed small communities which were leaning more towards internationalized collaborations. This might be groups mainly consisting of migrant scientists who collaborate with their former scientific organizations or they play a ``boundary spanning role'' among regional, national and continental contexts. We cannot investigate these type of questions at the organization level.

Furthermore, our definition of the Berlin metropolitan region was based on the affiliation addresses while literature on science geography presents a diverse array of definitions (e.g., Cottineau, Finance, Hatna, Arcaute, \& Batty, \protect\hyperlink{ref-cottineauDefiningUrbanAgglomerations2019}{2019}; Abbasiharofteh \& Broekel, \protect\hyperlink{ref-abbasiharoftehStillShadowWall2020}{2020}) from NUTS level to areas covering multiple cities which are overlooked in our data gathering strategy.

\hypertarget{acknowledgements}{%
\section{Acknowledgements}\label{acknowledgements}}

We would like to thank Sybille Hinze, Martin Reinhart and Paul Donner for comments and suggestions on earlier versions of this paper. Caveats are our responsibility.

\hypertarget{funding-information}{%
\section{Funding Information}\label{funding-information}}

This research was done in DEKiF project supported by Federal Ministry for Education and Research (BMBF), Germany, with grant number: M527600. Data is obtained from Kompetenzzentrum Bibliometrie (Competence Center for Bibliometrics), Germany, which is funded by BMBF with grant number 01PQ17001.

\hypertarget{data-availability}{%
\section{Data Availability}\label{data-availability}}

Data cannot be made publicly available due to the licensing and contract terms of the original data.

\hypertarget{references}{%
\section*{References}\label{references}}
\addcontentsline{toc}{section}{References}

\hypertarget{refs}{}
\leavevmode\hypertarget{ref-abbasiharoftehStillShadowWall2020}{}%
Abbasiharofteh, M., \& Broekel, T. (2020). Still in the shadow of the wall? The case of the Berlin biotechnology cluster: \emph{Environment and Planning A: Economy and Space}. \url{https://doi.org/10.1177/0308518X20933904}

\leavevmode\hypertarget{ref-akbaritabarConundrumResearchProductivity2018}{}%
Akbaritabar, A., Casnici, N., \& Squazzoni, F. (2018). The conundrum of research productivity: A study on sociologists in Italy. \emph{Scientometrics}, \emph{114}(3), 859--882. \url{https://doi.org/10.1007/s11192-017-2606-5}

\leavevmode\hypertarget{ref-akbaritabarGenderPatternsPublication2020}{}%
Akbaritabar, A., \& Squazzoni, F. (2020). Gender Patterns of Publication in Top Sociological Journals. \emph{Science, Technology, \& Human Values}. \url{https://doi.org/10.1177/0162243920941588}

\leavevmode\hypertarget{ref-akbaritabarItalianSociologistsCommunity2020}{}%
Akbaritabar, A., Traag, V. A., Caimo, A., \& Squazzoni, F. (2020). Italian sociologists: A community of disconnected groups. \emph{Scientometrics}. \url{https://doi.org/10.1007/s11192-020-03555-w}

\leavevmode\hypertarget{ref-amanHowCollaborationImpacts2016}{}%
Aman, V. (2016). How collaboration impacts citation flows within the German science system. \emph{Scientometrics}, \emph{109}(3), 2195--2216. \url{https://doi.org/10.1007/s11192-016-2092-1}

\leavevmode\hypertarget{ref-amanDoesScopusAuthor2018}{}%
Aman, V. (2018). Does the Scopus author ID suffice to track scientific international mobility? A case study based on Leibniz laureates. \emph{Scientometrics}, \emph{117}(2), 705--720. \url{https://doi.org/10.1007/s11192-018-2895-3}

\leavevmode\hypertarget{ref-araujoGenderDifferencesScientific2017}{}%
Araújo, E. B., Araújo, N. A. M., Moreira, A. A., Herrmann, H. J., \& Andrade, J. S. (2017). Gender differences in scientific collaborations: Women are more egalitarian than men. \emph{PLOS ONE}, \emph{12}(5), e0176791. \url{https://doi.org/10.1371/journal.pone.0176791}

\leavevmode\hypertarget{ref-avdeevInternationalCollaborationHigher2019}{}%
Avdeev, S. (2019). International Collaboration In Higher Education Research: A Gravity Model Approach. \emph{SSRN Electronic Journal}. \url{https://doi.org/10.2139/ssrn.3505886}

\leavevmode\hypertarget{ref-babchukCollaborationSociologyOther1999}{}%
Babchuk, N., Keith, B., \& Peters, G. (1999). Collaboration in sociology and other scientific disciplines: A comparative trend analysis of scholarship in the social, physical, and mathematical sciences. \emph{The American Sociologist}, \emph{30}(3), 5--21. \url{https://doi.org/10.1007/s12108-999-1007-5}

\leavevmode\hypertarget{ref-ballandComplexEconomicActivities2020}{}%
Balland, P.-A., Jara-Figueroa, C., Petralia, S. G., Steijn, M. P. A., Rigby, D. L., \& Hidalgo, C. A. (2020). Complex economic activities concentrate in large cities. \emph{Nature Human Behaviour}, \emph{4}(3), 248--254. \url{https://doi.org/10.1038/s41562-019-0803-3}

\leavevmode\hypertarget{ref-berlinuniversityallianceGemeinsamImVerbund2018}{}%
Berlin University Alliance. (2018, February). Gemeinsam im Verbund (Together as a group). \emph{Berlin University Alliance}. https://www.berlin-university-alliance.de/excellence-strategy/universities-of-excellence/index.html.

\leavevmode\hypertarget{ref-berlinuniversityallianceBerlinUniversityAlliance2019}{}%
Berlin University Alliance. (2019). \emph{Berlin University Alliance Proposal Crossing Boundaries toward an Integrated Research Environment}. Berlin: Berlin University Alliance.

\leavevmode\hypertarget{ref-biancaniSocialNetworksResearch2013}{}%
Biancani, S., \& McFarland, D. A. (2013). Social networks research in higher education. In \emph{Higher education: Handbook of theory and research} (pp. 151--215). Springer.

\leavevmode\hypertarget{ref-blumeSocialDirectionPublic1987}{}%
Blume, S., Bunders, J., Leydesdorff, L., \& Whitley, R. (Eds.). (1987). \emph{The Social Direction of the Public Sciences}. Dordrecht: Springer Netherlands. \url{https://doi.org/10.1007/978-94-009-3755-0}

\leavevmode\hypertarget{ref-breigerDualityPersonsGroups1974}{}%
Breiger, R. L. (1974). The Duality of Persons and Groups. \emph{Social Forces}, \emph{53}(2), 181--190. \url{https://doi.org/10.1093/sf/53.2.181}

\leavevmode\hypertarget{ref-cottineauDefiningUrbanAgglomerations2019}{}%
Cottineau, C., Finance, O., Hatna, E., Arcaute, E., \& Batty, M. (2019). Defining urban agglomerations to detect agglomeration economies. \emph{Environment and Planning B: Urban Analytics and City Science}, \emph{46}(9), 1611--1626. \url{https://doi.org/10.1177/2399808318755146}

\leavevmode\hypertarget{ref-dangeloCollectingLargescalePublication2020}{}%
D'Angelo, C. A., \& van Eck, N. J. (2020). Collecting large-scale publication data at the level of individual researchers: A practical proposal for author name disambiguation. \emph{Scientometrics}, \emph{123}(2), 883--907. \url{https://doi.org/10.1007/s11192-020-03410-y}

\leavevmode\hypertarget{ref-destefanoUseDifferentData2013}{}%
De Stefano, D., Fuccella, V., Vitale, M. P., \& Zaccarin, S. (2013). The use of different data sources in the analysis of co-authorship networks and scientific performance. \emph{Social Networks}, \emph{35}(3), 370--381. \url{https://doi.org/10.1016/j.socnet.2013.04.004}

\leavevmode\hypertarget{ref-donnerComparingInstitutionallevelBibliometric2019}{}%
Donner, P., Rimmert, C., \& van Eck, N. J. (2019). Comparing institutional-level bibliometric research performance indicator values based on different affiliation disambiguation systems. \emph{Quantitative Science Studies}, \emph{1}(1), 150--170. \url{https://doi.org/10.1162/qss_a_00013}

\leavevmode\hypertarget{ref-foxPublicationProductivityScientists1983}{}%
Fox, M. F. (1983). Publication Productivity among Scientists: A Critical Review. \emph{Social Studies of Science}, \emph{13}(2), 285--305. \url{https://doi.org/10.1177/\%2F030631283013002005}

\leavevmode\hypertarget{ref-glanzelBibliometricAnalysisInternational1999}{}%
Glänzel, W., Schubert, A., \& Czerwon, H. J. (1999). A bibliometric analysis of international scientific cooperation of the European Union (19851995). \emph{Scientometrics}, \emph{45}(2), 185--202. \url{https://doi.org/10.1007/BF02458432}

\leavevmode\hypertarget{ref-hoekmanResearchCollaborationDistance2010}{}%
Hoekman, J., Frenken, K., \& Tijssen, R. J. W. (2010). Research collaboration at a distance: Changing spatial patterns of scientific collaboration within Europe. \emph{Research Policy}, \emph{39}(5), 662--673. \url{https://doi.org/10.1016/j.respol.2010.01.012}

\leavevmode\hypertarget{ref-katzWhatResearchCollaboration1997}{}%
Katz, J., \& Martin, B. R. (1997). What is research collaboration? \emph{Research Policy}, \emph{26}(1), 1--18. \url{https://doi.org/10.1016/S0048-7333(96)00917-1}

\leavevmode\hypertarget{ref-katzGeographicalProximityScientific1994}{}%
Katz, J. S. (1994). Geographical proximity and scientific collaboration. \emph{Scientometrics}, \emph{31}(1), 31--43. \url{https://doi.org/10.1007/BF02018100}

\leavevmode\hypertarget{ref-laudelWhatWeMeasure2002a}{}%
Laudel, G. (2002). What do we measure by co-authorships? \emph{Research Evaluation}, \emph{11}(1), 3--15. \url{https://doi.org/10.3152/147154402781776961}

\leavevmode\hypertarget{ref-leaheySoleInvestigatorTeam2016}{}%
Leahey, E. (2016). From Sole Investigator to Team Scientist: Trends in the Practice and Study of Research Collaboration. \emph{Annual Review of Sociology}, \emph{42}(1), 81--100. \url{https://doi.org/10.1146/annurev-soc-081715-074219}

\leavevmode\hypertarget{ref-leonesciabolazzaDetectingAnalyzingResearch2017}{}%
Leone Sciabolazza, V., Vacca, R., Kennelly Okraku, T., \& McCarty, C. (2017). Detecting and analyzing research communities in longitudinal scientific networks. \emph{PLOS ONE}, \emph{12}(8), e0182516. \url{https://doi.org/10.1371/journal.pone.0182516}

\leavevmode\hypertarget{ref-luukkonenUnderstandingPatternsInternational1992}{}%
Luukkonen, T., Persson, O., \& Sivertsen, G. (1992). Understanding Patterns of International Scientific Collaboration. \emph{Science, Technology, \& Human Values}, \emph{17}(1), 101--126. \url{https://doi.org/10.1177/\%2F016224399201700106}

\leavevmode\hypertarget{ref-moedInternationalScientificCooperation1991}{}%
Moed, H. F., De Bruin, R. E., Nederhof, A. J., \& Tijssen, R. J. W. (1991). International scientific co-operation and awareness within the European community: Problems and perspectives. \emph{Scientometrics}, \emph{21}(3), 291--311. \url{https://doi.org/10.1007/BF02093972}

\leavevmode\hypertarget{ref-nederhofBibliometricMonitoringResearch2006}{}%
Nederhof, A. J. (2006). Bibliometric monitoring of research performance in the Social Sciences and the Humanities: A Review. \emph{Scientometrics}, \emph{66}(1), 81--100. \url{https://doi.org/10.1007/s11192-006-0007-2}

\leavevmode\hypertarget{ref-newmanScientificCollaborationNetworks2001}{}%
Newman, M. E. J. (2001a). Scientific collaboration networks. II. Shortest paths, weighted networks, and centrality. \emph{Physical Review E}, \emph{64}(1), 016132. \url{https://doi.org/10.1103/PhysRevE.64.016132}

\leavevmode\hypertarget{ref-newmanScientificCollaborationNetworks2001a}{}%
Newman, M. E. J. (2001b). Scientific collaboration networks. I. Network construction and fundamental results. \emph{Physical Review E}, \emph{64}(1), 016131. \url{https://doi.org/10.1103/PhysRevE.64.016131}

\leavevmode\hypertarget{ref-newmanDetectingCommunityStructure2004}{}%
Newman, M. E. J. (2004). Detecting community structure in networks. \emph{The European Physical Journal B - Condensed Matter}, \emph{38}(2), 321--330. \url{https://doi.org/10.1140/epjb/e2004-00124-y}

\leavevmode\hypertarget{ref-pallaQuantifyingSocialGroup2007}{}%
Palla, G., Barabási, A.-L., \& Vicsek, T. (2007). Quantifying social group evolution. \emph{Nature}, \emph{446}(7136), 664--667. \url{https://doi.org/10.1038/nature05670}

\leavevmode\hypertarget{ref-rammerKnowledgeProximityFirm2020}{}%
Rammer, C., Kinne, J., \& Blind, K. (2020). Knowledge proximity and firm innovation: A microgeographic analysis for Berlin. \emph{Urban Studies}, \emph{57}(5), 996--1014. \url{https://doi.org/10.1177/0042098018820241}

\leavevmode\hypertarget{ref-reichardtDetectingFuzzyCommunity2004}{}%
Reichardt, J., \& Bornholdt, S. (2004). Detecting fuzzy community structures in complex networks with a Potts model. \emph{Physical Review Letters}, \emph{93}(21), 218701. \url{https://doi.org/10.1103/PhysRevLett.93.218701}

\leavevmode\hypertarget{ref-rijckeEvaluationPracticesEffects2016}{}%
Rijcke, S. de, Wouters, P. F., Rushforth, A. D., Franssen, T. P., \& Hammarfelt, B. (2016). Evaluation practices and effects of indicator usea literature review. \emph{Research Evaluation}, \emph{25}(2), 161--169. \url{https://doi.org/10.1093/reseval/rvv038}

\leavevmode\hypertarget{ref-rimmertInstitutionalDisambiguationFurther2018}{}%
Rimmert, C. (2018). \emph{Institutional disambiguation for further countries - an exploration with extensive use of wikidata (project report).} (Report). Bielefeld: Bielefeld University, Institute for Interdisciplinary Studies of Science (I\(^2\)SoS).

\leavevmode\hypertarget{ref-smallSomeoneTalk2017}{}%
Small, M. L. (2017). \emph{Someone to Talk to}. Oxford University Press.

\leavevmode\hypertarget{ref-smallRoleSpaceFormation2019}{}%
Small, M. L., \& Adler, L. (2019). The Role of Space in the Formation of Social Ties. \emph{Annual Review of Sociology}, \emph{45}(1), 111--132. \url{https://doi.org/10.1146/annurev-soc-073018-022707}

\leavevmode\hypertarget{ref-smith-doerrHowDiversityMatters2017}{}%
Smith-Doerr, L., Alegria, S. N., \& Sacco, T. (2017). How Diversity Matters in the US Science and Engineering Workforce: A Critical Review Considering Integration in Teams, Fields, and Organizational Contexts. \emph{Engaging Science, Technology, and Society}, \emph{3}, 139. \url{https://doi.org/10.17351/ests2017.142}

\leavevmode\hypertarget{ref-sonnenwaldScientificCollaboration2007}{}%
Sonnenwald, D. H. (2007). Scientific collaboration. \emph{Annual Review of Information Science and Technology}, \emph{41}(1), 643--681. \url{https://doi.org/10.1002/aris.2007.1440410121}

\leavevmode\hypertarget{ref-stadtfeldMicroMacroLinkSocial2018}{}%
Stadtfeld, C. (2018). \emph{The Micro-Macro Link in Social Networks} (SSRN Scholarly Paper No. ID 3211795). Rochester, NY: Social Science Research Network.

\leavevmode\hypertarget{ref-stahlschmidtPerformanceStructuresGerman2019}{}%
Stahlschmidt, S., Stephen, D., \& Hinze, S. (2019). \emph{Performance and Structures of the German Science System} (p. 91). Studien zum deutschen Innovationssystem.

\leavevmode\hypertarget{ref-stephenPerformanceStructuresGerman2020}{}%
Stephen, D., Stahlschmidt, S., \& Hinze, S. (2020). \emph{Performance and Structures of the German Science System 2020}. Studien zum deutschen Innovationssystem.

\leavevmode\hypertarget{ref-subramanyamBibliometricStudiesResearch1983}{}%
Subramanyam, K. (1983). Bibliometric studies of research collaboration: A review. \emph{Journal of Information Science}, \emph{6}(1), 33--38. \url{https://doi.org/10.1177/016555158300600105}

\leavevmode\hypertarget{ref-traagNarrowScopeResolutionlimitfree2011}{}%
Traag, V. A., Van Dooren, P., \& Nesterov, Y. (2011). Narrow scope for resolution-limit-free community detection. \emph{Physical Review E}, \emph{84}(1), 016114. \url{https://doi.org/10.1103/PhysRevE.84.016114}

\leavevmode\hypertarget{ref-traagLouvainLeidenGuaranteeing2019}{}%
Traag, V. A., Waltman, L., \& van Eck, N. J. (2019). From Louvain to Leiden: Guaranteeing well-connected communities. \emph{Scientific Reports}, \emph{9}(1), 5233. \url{https://doi.org/10.1038/s41598-019-41695-z}

\leavevmode\hypertarget{ref-wagnerContinuingGrowthGlobal2015}{}%
Wagner, C. S., Park, H. W., \& Leydesdorff, L. (2015). The Continuing Growth of Global Cooperation Networks in Research: A Conundrum for National Governments. \emph{PLOS ONE}, \emph{10}(7), e0131816. \url{https://doi.org/10.1371/journal.pone.0131816}

\leavevmode\hypertarget{ref-wuchtyIncreasingDominanceTeams2007}{}%
Wuchty, S., Jones, B. F., \& Uzzi, B. (2007). The Increasing Dominance of Teams in Production of Knowledge. \emph{Science}, \emph{316}(5827), 1036--1039. \url{https://doi.org/10.1126/science.1136099}

\end{document}